\documentclass[twoside]{article}

\usepackage{float}
\usepackage{color}
\usepackage{greekbf}
\usepackage{float}
\usepackage{amsmath,amssymb}
\usepackage{mathrsfs}
\usepackage{enumerate}
\usepackage{srcltx}
\usepackage{mathtools}
    \mathtoolsset{showonlyrefs}
\usepackage{graphicx, psfrag}
\usepackage{setspace}
\usepackage{hyperref}
\hypersetup{bookmarks=false,
colorlinks=false,
linkcolor=black,
urlcolor=black,
pagebackref=true,
pdfstartview={FitH},
bookmarksopenlevel=2
}

\DeclareMathAlphabet{\mathbf}{OT1}{cmr}{bx}{it}

\definecolor{red}{rgb}{0.9,0,0}
\definecolor{blue}{rgb}{0.2,0.2,0.8}
\definecolor{green}{rgb}{0.0,0.5,0.2}
\definecolor{darkblue}{rgb}{0.2,0.2,0.5}
\definecolor{orange}{rgb}{1,0.5,0}
\definecolor{pink}{rgb}{0.96,0.5,0.46}
\definecolor{lblue}{rgb}{0.18,0.74,1}
\definecolor{cyan}{rgb}{0,0.8,0.8}

\newcommand {\xib}{\mathbf{\xi}}
\newcommand {\qb} {\mathbf{q}}

\newcommand {\chib}{\mathbf{\chi}}
\newcommand {\Bb} {\mathbf{B}}
\newcommand {\Ab} {\mathbf{A}}
\newcommand {\fb} {\mathbf{f}}

\newcommand {\Cc}  {\mathcal{C}}
\newcommand {\Dc}  {\mathcal{D}}
\newcommand {\Fc}  {\mathcal{F}}
\newcommand {\Hc}  {\mathcal{H}}
\newcommand {\Sc}  {\mathcal{S}}
\newcommand {\Tc}  {\mathcal{T}}
\newcommand {\Yc}  {\mathcal{Y}}

\newcommand {\db} {\mathbf{d}}

\begin{document}
	
\doublespacing

\title{\vspace{-3cm} {\bf  How graphene flexes and stretches \\under concomitant bending couples and tractions}}

\author{
Antonino Favata$^1$\!\!\!\!\! \and Andrea Micheletti$^2$\!\!\!\!\! \and Paolo Podio-Guidugli$^{3,4}$\!\!\!\!\! \and  Nicola M. Pugno$^{5,6,7}$
}

%

\maketitle

\vspace{-1cm}
\begin{center}
{\small

$^1$ Department of Structural and Geotechnical Engineering\\ Sapienza University of Rome, Italy\\[2pt]
\href{mailto: antonino.favata@uniroma1.it}{antonino.favata@uniroma1.it}\\[10pt]

$^2$ Dipartimento di Ingegneria Civile e Ingegneria Informatica\\
University of Rome TorVergata, Italy\\[2pt]
\href{mailto:micheletti@ing.uniroma2.it}{micheletti@ing.uniroma2.it}\\[8pt]
$^3$ Accademia Nazionale dei Lincei\\
Rome, Italy\\[5pt]
$^4$ Department of Mathematics\\University of Rome TorVergata, Italy \\[2pt] 
\href{mailto:ppg@uniroma2.it}{ppg@uniroma2.it}\\[8pt]
$^5$ Laboratory of Bioinspired and Graphene Nanomechanics\\ Department of Civil, Environmental and Mechanical Engineering\\ University of Trento, Italy\\[2pt]
\href{mailto:nicola.pugno@unitn.it}{nicola.pugno@unitn.it}\\[8pt]
$^6$ Center for Materials and Microsystems\\
Fondazione Bruno Kessler,  Trento, Italy\\[8pt]
$^7$ School of Engineering and Materials Science\\
Queen Mary University of London,  UK
}

\end{center}

\pagestyle{myheadings}
\markboth{A.~Favata, A.~Micheletti, P.~Podio-Guidugli, N.M.~Pugno}
{Graphene, bent and stretched}

\vspace{-0.5cm}
\section*{Abstract}

We propose a geometrically and physically nonlinear discrete mechanical model of graphene that assigns an energetic cost to changes in bond lengths, bond angles, and dihedral angles.  We formulate a variational equilibrium problem for a rectangular graphene sheet with assigned balanced forces and  couples uniformly distributed over opposite side pairs. We show that the resulting combination of stretching and bending
makes  achiral graphene easier to bend and harder (easier) to stretch  for small (large) traction loads. Our general developments hold for a wide class of REBO potentials; we  illustrate them in detail by numerical calculations performed in the case of a  widely used 2nd-generation Brenner potential.

\vspace{1cm}
\noindent {\large \bf Keywords}: graphene, softening, hardening, bending stiffness, stretching stiffness.

\tableofcontents

\section{Introduction}
\begin{figure}[h]\label{bending}
	\centering
	\includegraphics[scale=1.4]{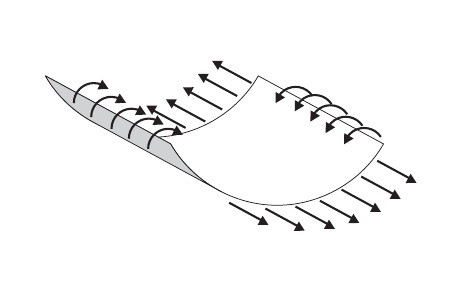}
	\caption{A sheet subject to  bending and stretching loads.}
\end{figure}

Flexible and stretchable  components are more and more frequently employed in such electronic devices as displays, light emitters, solar cells, etc. Recently, graphene has been shown to be a promising material to build devices that are able to bend and stretch; it has then become important to model, predict, and test, its mechanical behaviour under various combinations of  bending and stretching loads.
The basic problem we here tackle is sketched in Fig. \ref{bending}, where
a rectangular graphene sheet is depicted, subject to balanced couples and forces uniformly applied along  two different pairs of opposite sides. Our goal is to evaluate how the sheet's \textit{bending and stretching stiffnesses} depend on the value of the couple and force loads. A brief survey of the relevant literature is to be found in Section \ref{lit}.

Let us term the force distribution in Fig.~\ref{bending}  \emph{coherent} with the couple distribution. It has been shown in \cite{Shi_2012}, with the use of Density Functional Theory, that bending stiffness decreases in the presence of coherent traction loads. This effect, which is ascribed to a microscopic phenomenon, namely, the stretch-induced loosening of atom packing, is somehow counterintuitive for a person trained in standard structure mechanics, whose point of view is of course macroscopic.
In this contribution, we propose a \emph{geometrically and physically nonlinear discrete mechanical model}, microscopic in nature, which predicts, among other things, that coherent stretch reduces the bending stiffness of graphene.

Graphene is a two-dimensional carbon allotrope,  in the form of a hexagonal lattice whose vertices are occupied by C atoms.   In principle, just as for every other molecular arrangement, graphene's equilibrium shapes can be characterized as local minima of an intermolecular potential depending on the relative positions of \emph{all} C-atom pairs. Needless to say, numerical simulations based on such an approach would turn out to be either scarcely significant or prohibitively honerous. 
In another largely adopted approach, a choice of kinematic \emph{order parameters} is made: (i) under the assumption that their changes decide the energetic cost of shape changes, as predicted by a related intermolecular potential; (ii) with a view to end up with a list of Lagrangian coordinates much shorter than the collection of triplets of Cartesian coordinates of all C atoms in a given graphene sheet. 
What makes the list short is a careful account of the symmetries enjoyed by the problem at hand.
We choose three types of scalar order parameters, namely, bond length, bond angle, and dihedral angle (see Fig.~\ref{4toms} in Section \ref{riassunto}). The number of  triplets of such order parameters one needs assign to determine a deformed configuration of a graphene sheet is small when attention is confined, as we do here, to rectangular \emph{achiral graphene sheets} (see Fig.s  ~\ref{caseAA}  and \ref{caseZ}  for, respectively, armchair and zigzag graphene sheets), because their local and global geometries agree, in the sense that their chiral vectors  are parallel to a side pair, both before and after application of loads. A further reduction in the number of independent configuration variables follows from taking into account the symmetries in the equilibrium solutions implied by the specialty of the load distribution over the boundaries; as detailed in Sections \ref{eqbeA} and \ref{eqbeZ}, these symmetries are different in the two achiral cases, but equally effective.

Just as in \cite{Favata2015}, material constitution is specified here by an  intermolecular potential  depending on a finite list of the above order parameters: our constitutive prescriptions \eqref{V:AC}-\eqref{Va:AZ} are general enough to include all potentials in the REBO family we know of \cite{Brenner_1990,Brenner_2002,Tersoff_1988,Tersoff_1989}; equilibria correspond to local minima of an energy functional including also the potential of the applied loads; the governing equations are expressed in terms of three types of \textit{nanostresses}, that is, force-like objects which are work-conjugated to, respectively, changes in length of atomic bonds, changes in angle between two adjacent bonds, and changes in dihedral angles. The problem-specific novelties in this paper are that  we introduce proper definitions of curvature,  axial deformation, bending stiffness, and stretching stiffness of a graphene sheet; and that, for whatever potential in the chosen constitutive class, we derive two analytical conditions for, respectively, (i) softening of bending stiffness induced by a coherent force distribution (condition (C1),  formulated in Subsection \ref{softbend} for armchair graphene  and adapted for the zigzag case in Subsection \ref{5.3});\footnote{In this connection, we note that,  reversing the  force distribution shown in Fig.~\ref{bending} does not necessarily induce hardening, because the problem nonlinearity demands for a recalculation of the solution with \textit{a priori} unpredictable effetcs.} (ii) hardening of stretching stiffness induced by a coherent couple distribution (condition (C2), Subsections \ref{softyoung} and \ref{5.4}).   

Our main qualitative  result is that concomitant bending and stretching loads make the bending stiffness decrease and, provided the applied tractions are not too large, make the stretching stiffness increase; said differently, graphene is softer to bend when stretched and bent and harder to stretch when  bent and moderately stretched; moreover, the stretching stiffness decreases for large tractions, no matter how large the applied couples. Related qualitative results are the analytical expressions we derive for the \emph{pristine} ($\equiv$ no-load) bending and stretching stiffnesses of achiral graphene. These expressions permit to spot what deformation mechanisms
make graphene able to bear applied forces and couples, a piece of information that we regard as important to try and build a continuum theory by way of homogenization. 

To arrive at the representative quantitative results collected and discussed in Section \ref{rd}, we choose the same 2nd-generation Brenner potential as in \cite{Favata2015}.\footnote{In fact, we repeat, in addition to this one potential, our procedure is general enough to accommodate a variety of diehedral-angle sensitive REBO potentials; consequently, it can be adopted to find out whether an intermolecular potential in the class specified by \eqref{V:AC}-\eqref{Va:AZ} does predict the peculiar behavior of graphene predicted  in \cite{Shi_2012}.} We calculate numerically how bending stiffness (Fig.s \ref{ContD} - \ref{Dk}), stretching stiffness (Fig.s \ref{ContY} - \ref{Ye}) and nanostresses (Fig.s \ref{sigmaa} - \ref{T3}) depend on the axial strain and curvature induced by the applied force $F$ and couple $C$. As to the bending stiffness, we find that: (i) for $F=F^{\textrm{max}}$ and $C$ increasing from $0$ to $C^{\textrm{max}}$, it decreases till $\approx 19$\% ($43$\%) in the armchair (zigzag) case; (ii) for $F=0$ and $C$ increasing from $0$ to $C^{\textrm{max}}$, it decreases  till $\approx 20$\% ($23$\%) in the armchair (zigzag) case; (iii) $C=C^{\textrm{max}}$ and $F$ increasing from $0$ to $F^{\textrm{max}}$, it decreases till $\approx 35$\% ($59$\%) in the armchair (zigzag) case; (iv) for $C=0$ and $F$ increasing from $0$ to $F^{\textrm{max}}$,  it decreases  till $\approx 36$\% ($45$\%) in the armchair (zigzag) case.
As to the stretching stiffness, we find that,  for  $C$ increasing from $0$ to $C^{\textrm{max}}$, bending makes graphene harder to stretch when $F=0$ and  easier to stretch when $F=F^{\textrm{max}}$; the regime transition occurs at  the threshold value $F\simeq 6.5$ nN/nm in the armchair case ($F\simeq 12.3$ nN/nm in the zigzag case), that is, at about  15\% (29\%) of the fracture load. Moreover,  (i) for $F=F^{\textrm{max}}$ and $C$ increasing from $0$ to $C^{\textrm{max}}$, the stretching stiffness decreases till $\approx$ 14\% (13\%) in the  armchair (zigzag) case; (ii) for $F=0$ and $C$ increasing from $0$ to $C^{\textrm{max}}$, it increases till 
$\approx$ 11\% (38\%) in the armchair (zigzag) case.

\section{A brief survey of the literature}\label{lit}
The literature about the mechanical modeling of C-atom complexes whose shape is reminiscent of one or another type of macroscopic structure and whose dimensions are minuscules is vast: in addition to \emph{nanotubes}, by far the most studied such minuscule structures, one encounters \emph{nanoropes}, \emph{graphene nanoribbons}, \emph{nanoshells} (a term at times used as alternative to nanutubes), etc.; what follows has no pretensions to completeness, it only aims to exemplify the various modeling approaches that have been used.

The elastic properties of nanoropes and of single- and multi-wall carbon nanotubes (CNTs) have been investigated in \cite{LuJ1997} by means of a lattice model adopting a pair-wise harmonic interatomic potential; in \cite{Yakobson_1996}, molecular dynamics (MD) simulations were performed, on adopting the Tersoff-Brenner potential; local density approximation has been used in \cite{Tu_2002}, and \textit{ab initio} calculations relative to nanoshells are offered in \cite{Kudin_2001}; experimental values have been reported in \cite{Treacy_1996,Lourie_1998}.
Discrete models have been employed since long to predict the mechanical properties of 
CNTs:  in \cite{Geng2006},   closed-form expressions for the elastic properties of armchair  and zigzag  CNTs have been proposed; the model has been extended in \cite{Xiao2005} to study torsion loading, with  nonlinearities handled by means of a modified Morse potential.  A similar approach has been used in \cite{Shen_2004} to investigate various loading conditions, and in \cite{Wang2004} to evaluate  effective in-plane stiffness and bending rigidity of CNTs. In \cite{Chang_2005}, the model of \cite{Chang_2003} is extended to chiral  CNTs, an issue addressed also in \cite{Chang_2006}.  Computational methods have been presented in \cite{Meo2006} for CNTs and in \cite{Georgantzinos_2010,Geng2006,Georgantzinos_2011,Giannopoulos_2011,Giannopoulos_2012} for graphene and grahene nanoribbons.

Various continuum theories have been proposed, with the same scope: in \cite{Zhang_2002} a continuum theory of single-wall CNTs has been presented, based on Tersoff-Brenner potential; in \cite{Arroyo2004} the stretching and bending stiffnesses of graphene have been investigated, by means of Tersoff-Brenner interatomic potential; in \cite{Guo_2006,Wang_2006} the elastic properties of graphene and CNTs have been evaluated by means of a higher-order Cauchy-Born rule and of a Tersoff-Brenner potential; in \cite{Odegard_2002}, an equivalent-continuum modeling of nanostructured materials has been adopted;  and scale-bridging methods have been proposed in \cite{Chang_2010,Bajaj_2013,Favata2014a,Favata_2015b}.

As to single-layer graphene sheets, their nonlinear response to both in-plane and bending deformations  has been studied in \cite{Luhuang2009} and their out-of-plane bending behavior has been investigated in \cite{Scarpa_2009,Scarpa_2010} with the use a special equivalent atomistic-continuum model. In \cite{Zhou_2008}, the elastic properties of graphene have been theoretically predicted on taking into account internal lattice relaxation. Atomistic simulations have been employed to investigate the elastic properties of graphene in \cite{SakhaeePour_2009}.

To the best of our knowledge,   the effects on graphene stiffness of simultaneously increasing axial and bending loads have never been investigated by means of either first-principle calculations or MD. That graphene's bending stiffness decreases when graphene is stretched has been shown for the first time in 2012, in a paper where density functional theory (DFT) and bond-orbital-theory (BOT) calculations were performed  and found to provide coherent results \cite{Shi_2012}. A study of the bending stiffness of single-layer graphene with an initial curvature has been presented in \cite{Jomehzadeh_2013}, where small-scale effects are accounted for by the use of nonlocal elasticity. 	Very recently, in  \cite{Singh_2015} and in \cite{Singh_2015_a}, a continuum theory based on REBO potentials has been proposed, in order to study the elastic properties of finitely-deformed graphene sheets; the authors of \cite{Singh_2015}, seemingly unaware of \cite{Shi_2012}, claim that theirs is the first study where the effects of  curvature on stretching stiffness are considered. 

Papers \cite{Shi_2012} and \cite{Singh_2015} are the closest antecedents of our study, in which the effects of concomitant stretching and bending loads are examined in detail, both qualitatively and quantitatively, on the basis of the geometrically and physically nonlinear mechanical model we propose. In the first of those papers, the reported results are the output of  DFT and BOT computations; neither the dependence of the bending stiffness on the amount of bending is evaluated nor  the stretching stiffness is considered; the focus is on how bending stiffness depends on stretching in the absence of bending loads: it is shown that  softening does occur, a prediction of our model that we use to validate it by comparison. 
The influence of a modest amount of curvature on stretching stiffness has been considered in \cite{Singh_2015}, within the framework of a continuum theory and in the absence of stretching loads; a 2nd-generation Brenner potential is adopted, but dihedral contributions are neglected and the potential is not re-parameterized to compensate for this shortage. That dihedral contributions are important was pointed out in \cite{Lu_2009} and \cite{Favata2015}, and it is made evident by our equation \eqref{bendGS}, that displays a direct dependence of the bending stiffness, even in the absence of any loads, on the dihedral stiffness and the  bond-angle selfstress.

\section{A discrete mechanical model for graphene}\label{riassunto}
In this section, we introduce the geometrical, constitutive, and equilibrium-related, general features of our model. As anticipated, the solutions of the bending and stretching problems we solve will be discussed in detail in Section \ref{eqbeA} for an armchair sheet, and in Section \ref{eqbeZ}  for a zigzag  sheet.
\subsection{Order parameters}\label{KG}

\begin{figure}[h]
	\centering
	\includegraphics[scale=1]{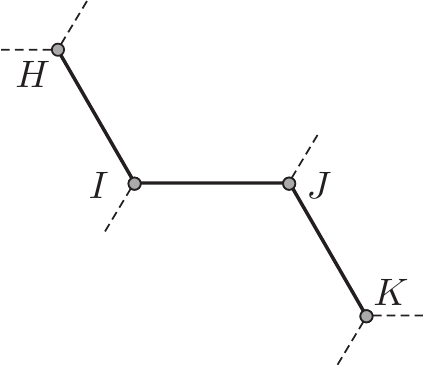}
	\caption{A four-atom chain in a hexagonal lattice.}
	\label{4toms}
\end{figure}

The kinematic variables we consider are associated with the interactions of a given atom with its first, second and third nearest neighbours. Precisely, with reference to the typical chain of four atoms depicted in Fig. \ref{4toms}, 
consisting of atom $H$ and its nearest neighbours $I$, $J$, and $K$, we consider: (i) \textit{bond lengths}, namely, the lengths of the covalent bonds between two successive atoms, such as $H$ and $I$; 
(ii) \textit{bond angles}, namely, the angles  between two successive bonds, such as $H-I$ and $I-J$;
(iii) \textit{dihedral  angles}, namely, the  angles  between the planes spanned by two pairs of successive bonds, such as the plane spanned by $H-I$ and $I-J$ and the plane spanned by $I-J$ and $J-K$.

\begin{figure}[h!]
	\centering
	\includegraphics[scale=1]{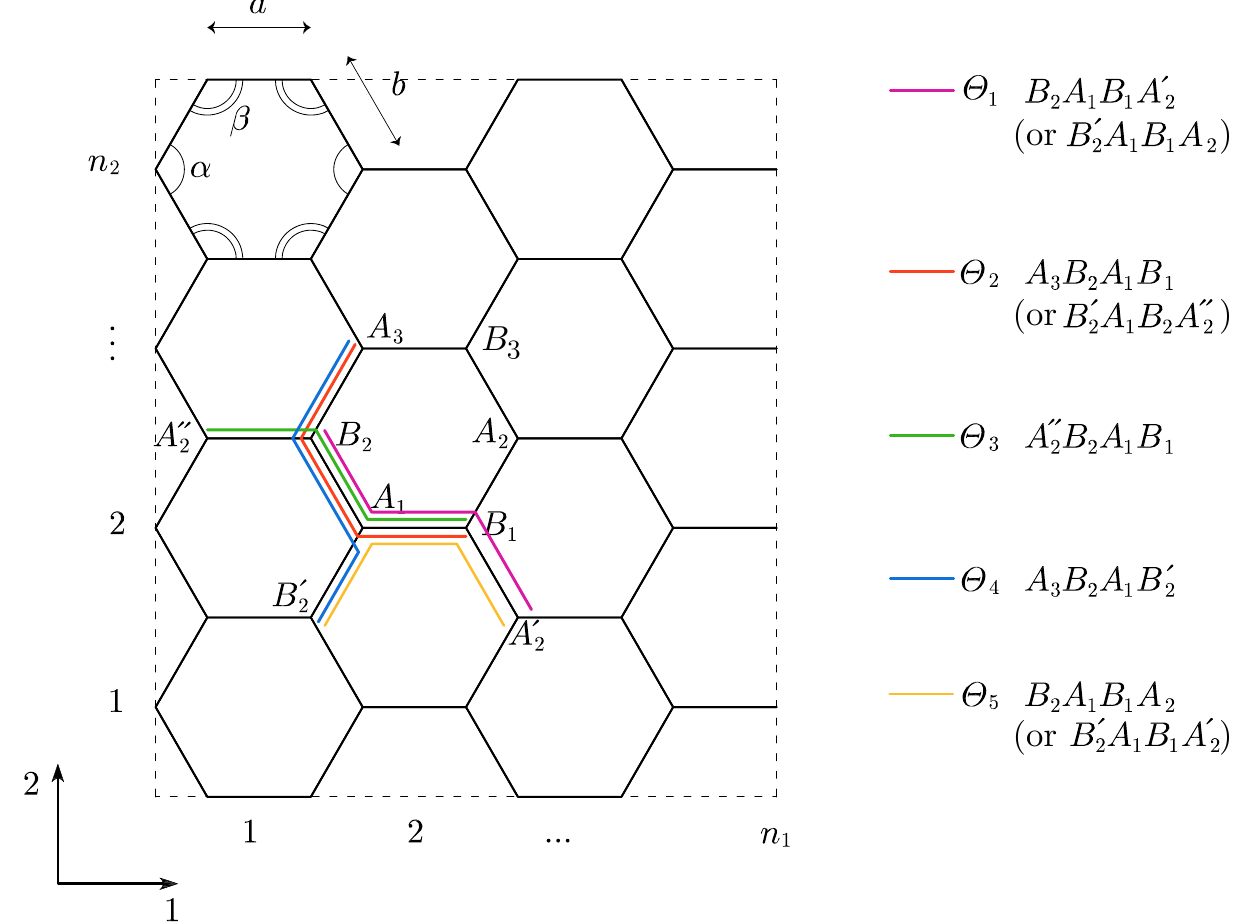}
	\caption{A rectangular graphene sheet.}
	\label{fig:AZ}
\end{figure}

In its unloaded ground configuration, the graphene sheet we consider has the form of a rectangle (see Fig. \ref{fig:AZ}), 
whose sides are aligned with, respectively, the armchair and zigzag directions $1$ and $2$; this rectangle consists of  $n_1$ hexagonal cells in direction $1$ and $n_2$ cells in direction $2$.
With reference to the representative cell $A_1 B_1 A_2 B_3 A_3 B_2 A_1$, $a$ denotes the length of sides $\overline{A_1 B_1}$ and $\overline{A_3 B_3}$, and $b$ the length of the remaining four sides; morever,
bond angles are of $\alpha$-type, such as e.g. $\widehat{A_3  B_2 A_1}$,  and of $\beta$-type, such as $\widehat{B_2 A_1 B_1}$; finally, there are only five types of dihedral angles, denoted by $(\Theta_1,\ldots,\Theta_5)$, to be individuated with the help  of  the colored bond chains.
The information carried by a 9-entry \emph{substring}
\begin{equation}\label{subs}
\xib_{sub}:=(a,b,\alpha,\beta,\Theta_1,\ldots,\Theta_5)
\end{equation}
is enough to determine the deformed configuration of a representative hexagonal cell. In principle, the whole sheet's \textit{order-parameter string} $\xib$ consists of an exhaustive and non-redundant sequence of cell substrings; in practice, as we shall see in Sections \ref{eqbeA} and \ref{eqbeZ},  the symmetries enjoyed by the equilibrium problems we  study are such that the convenient kinematical unknown turns out to be a string $\qb$ of Lagrangian coordinates shorter than $\xib_{sub}$. 
\subsection{Energetics and equilibria}\label{eande}
When attention is confined to coherent traction loads, a string $\qb$ of Lagrangian coordinates  is enough  to determine the energetic cost of  load-induced changes in bond lengths, bond angles and dihedral angles of achiral graphene sheets; this fact, that we anticipated in closing the previous subsection, will be made clear in Sections \ref{eqbeA} for armchair sheets and in Section  \ref{eqbeZ} for zigzag sheets.

The constitutive information about atom interactions is embodied in a a stored-energy functional  $V=\widetilde V(\xib)$ of the REBO class, to be specified in Section \ref{equilarm}; given the mapping $\qb\mapsto \xib=\widehat\xib(\qb)$, we set:
\begin{equation}\label{Vpot}
V=\widehat V(\qb),\quad\textrm{with}\quad \widehat V=\widetilde V\!\circ\widehat\xib\,.
\end{equation}
For $\fb$ the generalized \textit{dead load} work-conjugated to  the generalized displacement $\widehat{\db}(\qb)$, the \emph{load potential} takes the form:
\begin{equation}\label{lpot}
\widehat{P}(\qb):=\fb\cdot\widehat{\db}(\qb)\,.
\end{equation}
The \textit{equilibrium configurations} are the stationary points of the functional
\begin{equation}\label{totpot}
W=\widehat{W}(\qb):=\widehat{V}(\qb)-\widehat{P}(\qb)\,. 
\end{equation}
An equilibrium point $\qb_0$ satisfies the condition:
\begin{equation}\label{increE}
\delta W=\partial_\qb \widehat V(\qb_0)\cdot\delta\qb-\big(\partial_\qb\widehat{\db}(\qb_0) \big)^T\fb\cdot\delta\qb=0\quad \textrm{for all variations}\; \,\delta\qb=\qb-\qb_0,
\end{equation}
with
\[
\partial_\qb \widehat V(\qb_0)\cdot\delta\qb=\big(\partial_\qb \widehat\xib(\qb_0)\big)^T \partial_\xib \widetilde V(\xib_0)\cdot\delta\qb
=\partial_\xib \widetilde V(\xib_0)\cdot \big(\partial_\qb \widehat\xib(\qb_0)\big)\delta\qb\,,\quad \xib_0=\widehat\xib(\qb_0)\,.
\]
We set $\,\widetilde\chib:=\partial_\xib \widetilde V$, and call $\,\chib=\widetilde\chib(\xib)$ the \emph{stress mapping}, in that,
for $\,\delta\xib:=\xib-\xib_0$ the \emph{strain increment} in passing from the configuration $\xib_0$ to the configuration $\xib$, the quantity
\begin{equation}\label{increV}
\delta V=\chib\cdot\delta\xib
\end{equation}
can be regarded as the  \emph{incremental stress power}, that is, the expenditure of internal power associated with a strain increment. We also set $\,\widehat\Bb:=\partial_\qb\widehat\xib$,
and  call $\widehat\Bb$ the \emph{kinematic compatibility operator}, in that
\[
\delta\xib=\widehat\Bb(\qb)\delta\qb\,.
\]
Finally, we call $\widehat\Ab:=\widehat\Bb^T$ the \emph{equilibrium operator}, and note that \eqref{increE} holds if and only if
\begin{equation}\label{balc}
\widehat\Ab(\qb_0)\widetilde\chib(\xib_0)=\big(\partial_\qb\widehat{\db}(\qb_0) \big)^T\fb.
\end{equation}

\section{Armchair graphene}\label{eqbeA}
In the first two parts of this section we pose the equilibrium problem of an armchair graphene sheet acted upon by such couple and force distributions as depicted in Fig.~\ref{caseAA}. Our main concern is to assess whether coherent traction loads  induce a reduction in the bending  and in the stretching stiffnesses the graphene sheet exhibits in their absence {}; the preparatory developments to settle this issue in the affirmative in Section \ref{rd} are the contents of Subsection \ref{softbend}; the bending stiffness when traction loads are null is evaluated in Subsection \ref{softyoung}. 

The analysis in  Subsections \ref{lg} and \ref{equilarm} is general enough to evaluate the effects of any  combination of distributed couple and force loads; in particular, in the last subsection we determine the effects of couple loads on stretching stiffness and we evaluate the latter explicitly when those loads are null; that non-null couple loads induce hardening of stretching stiffness will be shown in Section \ref{rd}.
\subsection{Loads and geometry}\label{lg}
\begin{figure}[h]
	\centering
	\includegraphics[scale=1]{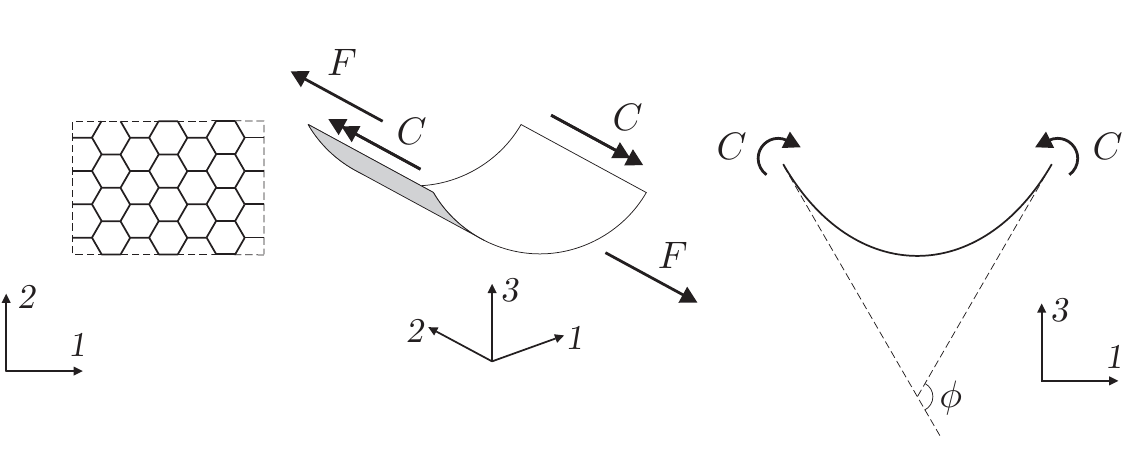}
	\caption{Bending and stretching an armchair graphene sheet.}
	\label{caseAA}
\end{figure}

The bending loads applied to the undeformed rectangular sheet sketched in the leftmost part of  Fig.~\ref{caseAA} consists of two equal and opposite sets of uniformly distributed couples, whose axes are aligned with direction $2$, applied over the two sides of the sheet parallel to direction $2$ itself;  $C$ is the magnitude of the resultant moment of both couple sets. Moreover, equal and opposite sets of uniformly distributed  forces parallel to direction $2$, whose resultant magnitude is $F$, are applied over the two sides of the rectangle parallel to direction $1$.\footnote{Couples and forces are {\em uniformly distributed} in a discrete sense.} 

The monolayer graphene piece we consider is in its \emph{ground configuration} (GC) when both $C$ and $F$ are null, all atoms lie on the same plane, all bond lengths have the ground length $r_0$,\footnote{The value of this parameter depends slightly on the intermolecular potential of one's choice; for the 2nd-generation Brenner potential we use later on in our computations, $r_0=0.14204$ nm.} all bond angles are equal to $\theta_0=2\pi/3$, and all dihedral angles are null; we assume that the stored-energy functional $\widetilde V$ has a global minimum in the GC. 
When at least one of $C$ and $F$ is non-null, the graphene piece is in a \emph{deformed configuration} (DC). 
No matter if $F$ is null or not, when $C\neq 0$ we have that: (i) all  atoms lie on the lateral surface of a right cylinder, whose axis is parallel to direction $2$; (ii) each plane orthogonal to the cylinder's axis and passing through an atom is a plane of both reflection and periodic symmetry for atomic positions; (iii) the axis of the cylinder is an axis of one-cell periodic polar symmetry for atomic positions, and any plane through this axis and an atom is a plane of reflection symmetry.
Consequently, in a DC there are only two inequivalent bond length $a$ and $b$, two inequivalent bond angles $\alpha$ and $\beta$, and three inequivalent dihedral angles, $\Theta_i$ $(i=1,2,3)$, while the two remaining dihedral angles keep the GC value $\Theta_0=0$.

\begin{figure}[h]
	\centering
	\includegraphics[scale=0.8]{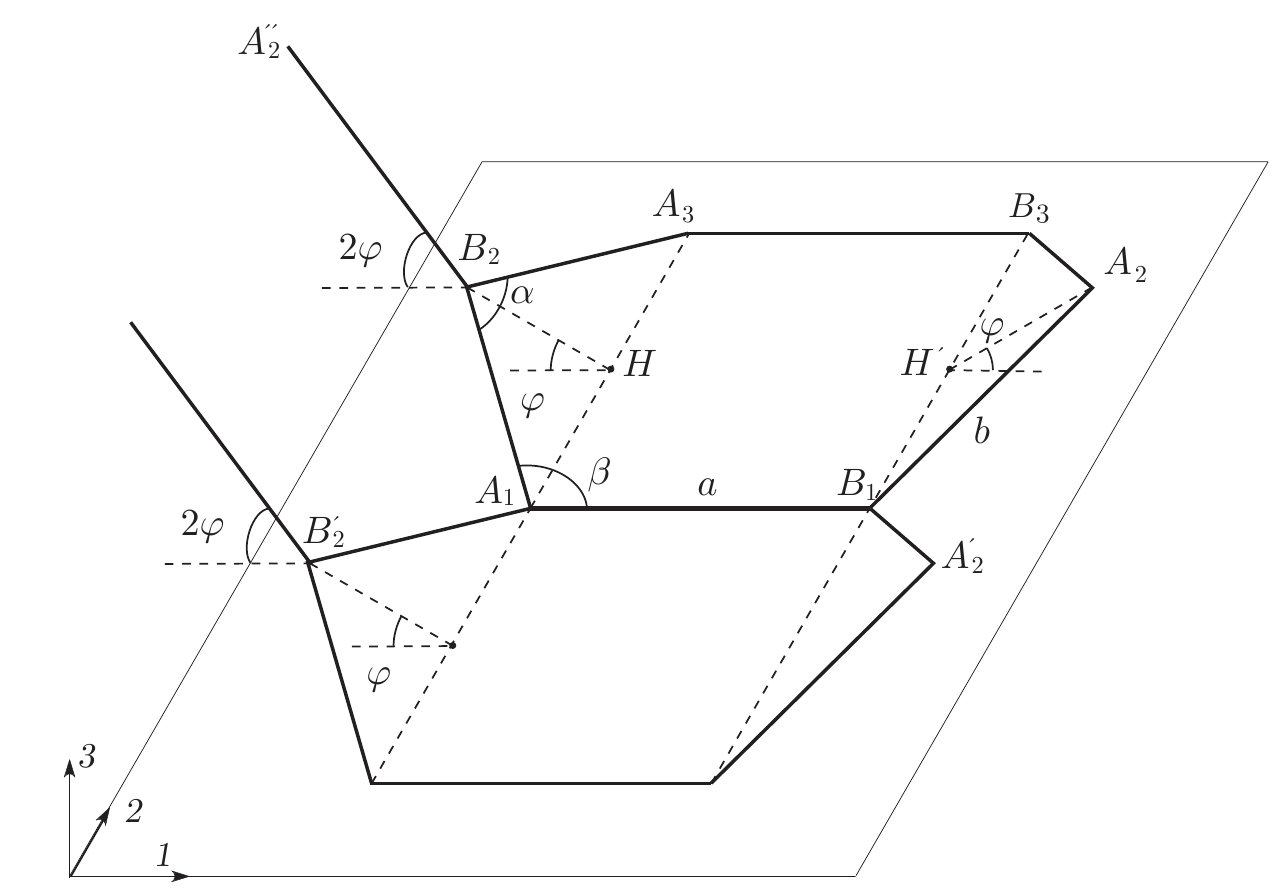}
	\caption{The deformed cell of an armchair graphene sheet.}
	\label{cell}
\end{figure}

With reference to Fig. \ref{cell},
let $\varphi$  be the angle between the plane of $A_1, B_1$, and $B_3$ and the plane of $B_1, A_2$, and $B_3$ and, this time with reference to the rightmost sketch in Fig.~\ref{caseAA}, let  $\Phi$ be the angle at center subtended by the deformed sheet; then,
\begin{equation}\label{eqn2}
\Phi=2n_1\varphi\,.
\end{equation}
Moreover, for geometric compatibility, the bond angles $\alpha$ and $\beta$ must satisfy the following condition:
%
%
\begin{equation}\label{geomcompA}
\cos\beta=-\cos\frac{\alpha}{2}\cos\varphi\,,
\end{equation}
whence
\begin{equation}\label{betafunA}
\beta=\widetilde{\beta}(\alpha,\varphi):=\arccos\left(-\cos\frac{\alpha}{2}\cos\varphi\right).
\end{equation}
Finally, the dihedral angles can be expressed in terms of $\alpha$ and $\beta$ with the use of the following relations:
\begin{equation} \label{gamma1A}
\sin\beta\,\sin\frac{\Theta_1}{2}=\cos\frac{\alpha}{2}\sin\varphi\,,\quad
\sin\beta\,\sin\Theta_2=\sin\varphi\,, \quad
\Theta_3=2\,\Theta_2\,, \quad \Theta_4= \Theta_5=0\,,
\end{equation}
whence expressions for $\Theta_1=\widetilde\Theta_1(\alpha,\varphi)$ and $\Theta_2=\widetilde\Theta_2(\alpha,\varphi)$ follow.

With the help of Figure~\ref{radii}, it is not difficult to see that the following geometric compatibility relation holds:
\begin{figure}[h!]
	\centering
	\includegraphics[scale=0.86]{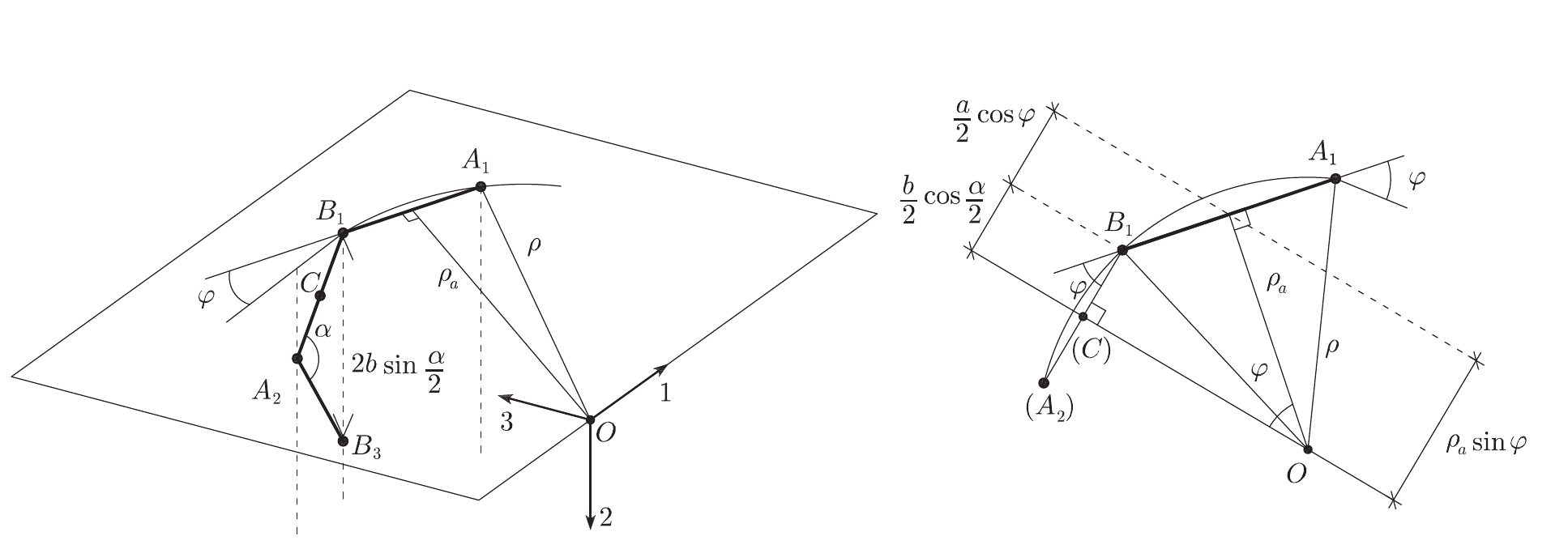}
	\caption{Local geometry of a deformed armchair graphene sheet.}\label{radii}
\end{figure}
\begin{equation}\label{rhoa}
\rho_a\sin\varphi=\frac{b}{2}\cos\frac{\alpha}{2}+\frac{a}{2}\cos\varphi\,,
\end{equation}
where $\rho_a$ is the distance of an $a$-type bond from the cylinder's axis; moreover,
\begin{equation}\label{RA}
\rho=\sqrt{\rho_a^2+\frac{a^2}{4}}\,,
\end{equation}
where $\rho$ denotes the cylinder's radius; hence, the \textit{current curvature} $\kappa:=1/\rho$ has the expression
\begin{equation}\label{kappa}
\kappa=\left(\rho_a^2+\frac{a^2}{4}\right)^{-1/2},\quad \rho_a=(\sin\varphi)^{-1}\left(\frac{b}{2}\cos\frac{\alpha}{2}+\frac{a}{2}\cos\varphi\right).
\end{equation}
Moreover, the \emph{current lengths} of the rectangle's sides are given by:
\begin{equation}\label{L2A}
\begin{aligned}
\lambda_1&=\left(1-\frac{b}{a}\cos\beta \right)n_1\,a,\;\;\textrm{in direction}\;\,1,
\\
\lambda_2&=2\,n_2\, b\,\sin\frac{\alpha}{2},\;\;\textrm{in direction} \;\,2;
\end{aligned}
\end{equation}
as an axial deformation measure we take
\begin{equation}\label{strain}
\varepsilon=\frac{\lambda_2-\lambda_{2,0}}{\lambda_{2,0}},\quad \lambda_{2,0}=\sqrt{3}\,n_2r_0,
\end{equation}
where $\lambda_{2,0}$ denotes the GC length of the rectangle's side in direction 2, whence
\begin{equation}\label{epsi}
\varepsilon=\frac{2}{\sqrt 3}\frac{b}{r_0}\,\sin\frac{\alpha}{2}-1\,.
\end{equation}

\subsection{Equilibrium conditions}\label{equilarm}
Let $n_a, n_b, n_\alpha, n_\beta$, and $n_{\Theta i}$ $(i=1,2,3)$, be the number of bond lengths, bond angles, and dihedral angles, of the same type; it is the matter of a simple count to find:
\begin{equation}\label{ns}
\begin{aligned}
& n_a \simeq n_1 n_2,\quad n_b = 2 n_1 n_2,\quad n_\alpha \simeq 2 n_1 n_2,\quad n_\beta = 4 n_1 n_2,\\
& n_{\Theta_1} = 2 n_a,\quad n_{\Theta_2} = 2n_b,\quad n_{\Theta_3} =  n_b, \quad \quad n_{\Theta_4}=n_b, \quad n_{\Theta_5}=2n_a
\end{aligned}
\end{equation}
(in the above expression, terms which are linear in $n_1$ or $n_2$, two very large integers, have been neglected).

Firstly, we specify as follows the stored-energy functional introduced in equation \eqref{Vpot}:
\begin{equation}\label{V:AC}
V=n_a V_a + n_b V_b = n_1 n_2 (V_a+2V_b)\,,
\end{equation}
where
\begin{equation}\label{Va:AZ}
\begin{aligned}
V_a(a,\beta,\Theta_1)&=V_R(a)+b_a(\beta, \Theta_1)\,V_A(a)\,,\\
V_b(b,\alpha,\beta,\Theta_2,\Theta_3,\Theta_4)&=V_R(b) +
b_b(\alpha, \beta, \Theta_2, \Theta_3, \Theta_4)\,V_A(b)\,.
\end{aligned}
\end{equation}
As is typical of REBO potentials, $V_R$ and $V_A$ are, respectively, the repulsive and attractive ingredients of $V$,  and $b_a, b_b$ are the bond-order mappings, each of which depends on some, but not all, of the geometric parameters. In the present case of armchair bending, with slight notational abuse, we set:
\begin{equation}\label{achipot}
V=\widetilde V(\xib):=n_aV_a(a,\beta,\Theta_1)+n_bV_b(b,\alpha,\beta,\Theta_2,\Theta_3,0),\quad\xib:=(a,b,\alpha,\beta,\Theta_1,\Theta_2,\Theta_3,0,0);\footnote{Here we have taken relations $\eqref{gamma1A}_{4,5}$ into account;  later on, when we deal with zigzag bending, we shall use  another specialization of \eqref{V:AC} and \eqref{Va:AZ}.}
\end{equation}
due to the geometric conditions 
\eqref{geomcompA} and \eqref{gamma1A}, the 9-entry string $\xib_{sub}$ in \eqref{subs} is determined by the 4-entry string $\qb=(a,b,\alpha,\varphi)$ (recall the anticipations given in Sections \ref{KG} and \ref{eande}).

Secondly, we specify as follows the load potential introduced in equation \eqref{lpot}:
\begin{equation}
P=F(\lambda-\lambda_0)+C\,\Phi,
\end{equation}
where the loads $\fb\equiv(F,C),$ are associated to the generalized displacement
\[
\begin{aligned}
\widehat\db(\qb)&\equiv \big(\delta\widehat\lambda(b,\alpha),\delta\widehat\Phi(\varphi)\big),
\\
\delta\widehat\lambda(b,\alpha)&=\lambda-\lambda_0=2n_2\, b\,\sin\frac{\alpha}{2}-\lambda_0,\quad \delta\widehat{\Phi}(\varphi)=\Phi=2n_1\varphi
\end{aligned}
\]
(here we have made use of \eqref{L2A} and \eqref{eqn2}). 

We are now in a position to write explicitly the stationarity condition \eqref{balc} of  potential $\widehat W$. Under the present circumstances,  the equilibrium operator takes the form of a $4\times 7$ matrix:
\begin{equation}\label{A}
\left[\begin{array}{c}\Ab\end{array}\right]=\left[\begin{array}{cccccccc}
1 & 0 & 0 & 0 & 0 & 0 & 0  \\
0 & 1 & 0 & 0 & 0 & 0 & 0  \\
0 & 0 & 1 & \beta,_\alpha & \Theta_1,_\alpha & \Theta_2,_\alpha & 2\Theta_2,_\alpha \\
0 & 0 & 0 & \beta,_\varphi & \Theta_1,_\varphi & \Theta_2,_\varphi & 2\Theta_2,_\varphi 
\end{array}\right]\,,
\end{equation}
and the stress-mapping string $\chib$ consists of the following seven entries:
\begin{equation}\label{chi}
\left[\,\chib\,\right]^T=\left[\,
n_a\,\sigma_a \quad
n_b\,\sigma_b  \quad
n_\alpha\,\tau_\alpha \quad
n_\beta\,\tau_\beta\quad
n_{\Theta_1} \, \Tc_1\quad
n_{\Theta_2}  \, \Tc_2\quad
n_{\Theta_3}   \, \Tc_3
\right]\,.
\end{equation}
All in all, in view of \eqref{ns} and $\eqref{gamma1A}_{3}$, the equilibrium equations \eqref{balc} now read:
\begin{eqnarray}
& &\sigma_a= 0\,,\label{f1:A3}\\
& &\sigma_b - \frac{F}{n_1}\sin\frac{\alpha}{2}= 0\,,\label{f2:A3}\\
& &\tau_\alpha+2\tau_\beta \beta,_\alpha+\Theta_1,_\alpha\Tc_1+2\Theta_2,_\alpha(\Tc_2+\Tc_3)- \frac{1}{2}\frac{F}{n_1}b\cos\frac{\alpha}{2}= 0\,,\label{f3:A3}\\
& &2\tau_\beta \beta,_\varphi+\Theta_1,_\varphi\Tc_1+2\Theta_2,_\varphi(\Tc_2+\Tc_3)-\frac{C}{n_2} = 0\,.\label{f4:A3}
\end{eqnarray}
We call each  of $\sigma_a, \sigma_b, \tau_\alpha, \tau_\beta$, and $\Tc_i$ $(i=1,2,3)$, a {\em nanostress}. Nanostresses are work-conjugate to changes of, respectively,  $a$- and $b$-bond lengths, $\alpha$ and $\beta$ bond angles, and dihedral angles, and depend as follows from the order parameters:
\begin{eqnarray}
&& \sigma_a  = V_R'(a)+b_a(\beta,\Theta_1)\,V_A'(a)\,,\quad
\sigma_b  = V_R'(b)+b_b(\alpha,\beta,\Theta_2,2\Theta_2,0)\,V_A'(b)\,, \label{nanostress1}\\
&& \tau_\alpha  = b_b,_\alpha\!(\alpha,\beta,\Theta_2,2\Theta_2,0)\, V_A(b) \,,\quad
\tau_\beta  = \frac{1}{4}\big(b_a,_\beta\!(\beta,\Theta_1) \,V_A(a)+ 2 b_b,_\beta\!(\alpha,\beta,\Theta_2,2\Theta_2,0) \,V_A(b)\big)\,,\label{nanostress2}
\\
&& \hspace{-0.75cm}\Tc_1  = \frac{1}{2}\,b_a,_{\Theta_1}\!(\beta,\Theta_1) \,V_A(a)\,,\,
\Tc_2  =  \frac{1}{2}\,b_b,_{\Theta_2}\!(\alpha,\beta,\Theta_2,2\Theta_2,0) \,V_A(b)\,,\,  \Tc_3= \frac{1}{2}\,b_b,_{2\Theta_2}\!(\alpha,\beta,\Theta_2,2\Theta_2,0) \,V_A(b)\label{nanostress3}
\end{eqnarray}
(in the first two of these relations an apex signifies differentiation). Combination of \eqref{nanostress1}-\eqref{nanostress3} with \eqref{betafunA} and $\eqref{gamma1A}_{1,2}$, followed by substitution into \eqref{f1:A3}$-$\eqref{f4:A3}, gives a nonlinear system of four equations in the four unknowns $(a,b,\alpha,\varphi)$, whose solution is unique under reasonable assumptions on the  constitutive mappings $V_R,V_A,b_a$ and $b_b$,\footnote{For an example of such assumptions, which are fulfilled by the stored-energy functional we will use to obtain the representative results reported in Section \ref{rd},  see \cite{Favata2015}.} a system generally too difficult to be solved analytically but solvable numerically for any given load pair $(C,F)$.

\subsection{Stretching-induced softening of bending stiffness}\label{softbend}
In the framework of structure mechanics,  bending stiffness is a constitutive/geometric notion, intended to measure a structure's sensitivity to bending \emph{whatever the loads}: e.g., in the standard one-dimensional linear theory of rods, the bending stiffness $EJ$, where $E$ is the Young's modulus and $J$ the cross-section's inertia, equals the moment-to-curvature ratio $M/\kappa$; its inverse $(EJ)^{-1}$ is the bending compliance. In the present context, the role of stiffness or compliance notions is different: although they all  incorporate constitutive and geometric information, they tell us about \emph{a structure's response to a given system of loads} and are expressed in terms of the solution to the relative equilibrium problem. Had we at our disposal an analytic expression for the part $\varphi=\widehat\varphi(C,F)$ of the solution to the armchair-bending problem, our task would be easy: in view of \eqref{eqn2}, we would set
\[
\Phi=\widehat\Phi(C,F):=2n_1\widehat\varphi(C,F)
\]
and define the bending compliance of a graphene sheet to be 
\[
\widehat\gamma(C,F):=\partial_C\widehat\Phi(C,F),
\]
with
\[
\partial_F\widehat\gamma(C,F):=\partial_{FC}^{(2)}\widehat\Phi(C,F)
\]
the relative stretching sensitivity:
the sign of this second derivative would tell us whether an axial traction induces softening or hardening of the sheet's bending stiffness. Unfortunately, we cannot count on an explicit knowledge of $\widehat\varphi$. We then take a different and less direct path.

When combined with \eqref{nanostress1}-\eqref{nanostress3}  and \eqref{L2A}$_2$, equation \eqref{f4:A3} can be used to define a mapping $\widetilde\Cc$ delivering the couple per  unit current length:
\begin{equation}\label{coppia}
\Cc=\widetilde\Cc(a,b,\alpha,\beta,\Theta_1,\Theta_2):=\left( 2b\sin\frac{\alpha}{2} \right)^{-1}\big(2\tau_\beta \beta,_\varphi+\Tc_1\Theta_1,_\varphi+2(\Tc_2+\Tc_3)\Theta_2,_\varphi\big).
\end{equation}
Now, in view of the geometric relations \eqref{betafunA} and $\eqref{gamma1A}_{1,2}$, each of the variables $\beta$, $\Theta_1$ and $\Theta_2$  depends in a known manner on the pair $(\alpha,\varphi)$. Consequently,  (i) in view of the equilibrium equation \eqref{f1:A3} and the constitutive equation  $\eqref{nanostress1}_1$, variable $a$ too depends in a known manner on $(\alpha,\varphi)$; (ii) in view of the equilibrium equation \eqref{f2:A3} and the constitutive equation \eqref{nanostress1}$\eqref{nanostress1}_2$, variable $b$ depends in a known manner on $(\alpha,\varphi)$ \emph{and} the datum $F$: we provisionally have from \eqref{coppia} that
\begin{equation}\label{coppiab}
\Cc=\widetilde\Cc\big(\widetilde a(\alpha,\varphi),\widetilde b(\alpha,\varphi,F),\alpha,\widetilde\beta(\alpha,\varphi),\widetilde\Theta_1(\alpha,\varphi),\widetilde\Theta_2(\alpha,\varphi)\big).
\end{equation}
Furthermore,  in view of the dependences detailed just above, (iii) when combined with the constitutive equations $\eqref{nanostress2}$, the equilibrium equation \eqref{f3:A3} takes the form of a restriction on the triplet $(\alpha,\varphi, F)$; (iv) the geometric relation \eqref{kappa} takes the form of a restriction on the triplet $(\alpha,\varphi, \kappa)$, where, we recall, $\kappa$ is the current curvature; (v) the system of these two restrictions provides implicit representations in terms of the pair $(\kappa,F)$ for both $\alpha$ and $\varphi$.
In conclusion, we have from \eqref{coppiab} that
\begin{equation}\label{coppiac}
\Cc=\widehat\Cc(\kappa,F):=\widetilde\Cc\big(\widetilde a(\alpha,\varphi),\widetilde b(\alpha,\varphi,F),\alpha,\widetilde\beta(\alpha,\varphi),\widetilde\Theta_1(\alpha,\varphi),\widetilde\Theta_2(\alpha,\varphi)\big), \quad
\alpha=\widetilde\alpha(\kappa,F),\;\, \varphi=\widetilde\varphi(\kappa,F).
\end{equation}

For the \textit{bending stiffness} of an armchair graphene sheet loaded as indicated in Fig. \ref{caseAA} we take 
\begin{equation}\label{bendsti}
\Dc=\widehat\Dc(\kappa,F):=\partial_\kappa\widehat\Cc(\kappa,F);
\end{equation}
the derivative of $\widehat\Dc$ with respect to $F$:
\begin{equation}\label{softening}
\Sc=\widehat\Sc(\kappa,F):=\partial_F\widehat\Dc(\kappa,F)=\partial_{F\kappa}^{(2)}\widehat\Cc(\kappa,F)
\end{equation}
measures the sensitivity of the bending stiffness to the applied traction: 
\begin{itemize}
	\item[(C1)] \emph{stretching-induced softening of the bending stiffness occurs whenever} $\Sc<0$.
\end{itemize}
We point out that these derivatives can be analytically determined, possibly with the help of a symbolic manipulator, but the task of evaluating them in correspondence of the solution of a given equilibrium problem can be achieved only numerically; this we have done, and our findings are presented in Sec. \ref{5.1}. To evaluate the pristine bending stiffness of an armchair graphene sheet -- that is, its bending stiffness for null force loads and small angle $\varphi$ -- turns out to be a much easier task, undertaken in the next section.

\subsection{Pristine bending stiffness}\label{pristbend}
We consider an armchair graphene sheet in its ground configuration (GC). In addition to setting $F=0$ in the equilibrium equations \eqref{f2:A3} and \eqref{f3:A3}, we approximate all $\varphi$-dependences detailed just after equation \eqref{coppia} by their first-order expansion in $\varphi$ itself; combination of these two measures eliminates any need for numerical calculations. Our developments are sketched hereafter.

To begin with, note that the first-order expansion in $\varphi$ of \eqref{kappa} is:
\begin{equation}
\kappa_0:=\frac{4}{3}\,{r_0}^{-1}\varphi.
\end{equation}
In view of \eqref{coppia} and \eqref{bendsti}, we have that, respectively,
\begin{equation}\label{coppia0}
\Cc_0=(\sqrt{3}\,r_0)^{-1}\big(2\tau_\beta \beta,_\varphi+\Tc_1\Theta_1,_\varphi+2(\Tc_2+\Tc_3)\Theta_2,_\varphi\big)\Big|_{\rm GC}
\end{equation}
and 
\begin{equation}\label{bendsti0}
\Dc_0:=\left(\varphi,_\kappa  \partial_\varphi\Cc  \right)\Big|_{\rm GC}=\frac{\sqrt{3}}{4}\,\partial_\varphi \big(2\tau_\beta \beta,_\varphi+\Tc_1\Theta_1,_\varphi+2(\Tc_2+\Tc_3)\Theta_2,_\varphi\big)\Big|_{\rm GC},
\end{equation}
where the indicated evaluations at the ground configuration are understood in the sense of small angles $\varphi$. Recourse to the constitutive equations \eqref{nanostress1}-\eqref{nanostress3}, after some simple calculations that we here omit, yields the following formula:
\begin{equation}\label{bendGS}
\Dc_0=-\frac{1}{2}\,\tau_0+\frac{7}{\sqrt{3}}\,\nu,
\end{equation}
where 
\[
\tau_0:=\tau_\alpha\Big|_{\rm GC}=\tau_\beta\Big|_{\rm GC} 
\] 
is the \emph{bond-angle self-stress} -- that is, the ground value of the nanostress work-conjugated to bond-angle changes -- and $\nu$ is the \emph{dihedral stiffness}  in GC -- that is, the ground value of the nanostress work-conjugated to a unit change in dihedral angle:
$$
\Tc_i\Big|_{\rm GC}=\nu\,\Theta_i\Big|_{\rm GC},   
\;\, (i=1,2,3).
$$		

When, as we do in Sec. \eqref{rd}, we specialize the above results for the case of a 2nd-generation Brenner potential, we find that  $\,\nu $= 0.0282596 nN$\times$nm; moreover, we import from  \cite{Favata2015} the value $\tau_0 = -0.2209$ nN$\times$nm for the bond-angle self-stress. The resulting GC bending stiffness has a value in complete agreement with  what is found in the literature \cite{Lu_2009}, namely,
\begin{equation}\label{bs}\Dc_0=0.22466\; {\rm nN}\!\times\!{\rm nm}=1.4022\;{\rm eV}.
\end{equation}
We point out that, in absence of the dihedral contribution, the bending stiffness is about one half of the above value. Such an incorrect evaluation is inevitable, as remarked in \cite{Lu_2009}, in case of MD simulations based on potentials that do not account for third-neighbour interactions (e.g., the 1st-generation Brenner potential).

\subsection{Bending-induced hardening of stretching stiffness}\label{softyoung}
In this subsection we parallel as much as we can the developments in the previous one. Just as we did to lay down equation \eqref{coppia} for the couple per unit current  length, we begin by observing that,
when combined with \eqref{nanostress2}-\eqref{nanostress3} and \eqref{L2A}, equation \eqref{f2:A3} can be used to define a mapping $\widetilde\Fc$ delivering the force per  unit current length:
\begin{equation}\label{forza}
\Fc=\widetilde\Fc(a,b,\alpha,\beta,\Theta_1,\Theta_2):=\sigma_b\left( \left(1-\frac{b}{a}\cos\beta \right)\sin\frac{\alpha}{2} \right)^{-1}
\end{equation}
We continue by noting that,
in view of the geometric relations \eqref{betafunA} and $\eqref{gamma1A}_{1,2}$, each of the variables $\beta$, $\Theta_1$ and $\Theta_2$  depends in a known manner on the pair $(\alpha,\varphi)$. Consequently,  (i) in view of the equilibrium equation \eqref{f1:A3} and the constitutive equation  $\eqref{nanostress1}_1$, variable $a$ too depends in a known manner on $(\alpha,\varphi)$; (ii) in view of the equilibrium equation \eqref{f3:A3} and the constitutive equation $\eqref{nanostress1}_2$, variable $b$ depends in a known manner on $(\alpha,\varphi)$ \emph{and} the datum $C$. Thus, we provisionally have from \eqref{forza} that
\begin{equation}\label{forzab}
\Fc=\widetilde\Fc\big(\widetilde a(\alpha,\varphi),\widetilde b(\alpha,\varphi,C),\alpha,\widetilde\beta(\alpha,\varphi),\widetilde\Theta_1(\alpha,\varphi),\widetilde\Theta_2(\alpha,\varphi)\big).
\end{equation}
Furthermore,  in view of the dependences detailed just above, (iii) the equilibrium equation \eqref{f3:A3}, when combined with the constitutive equations \eqref{nanostress2} and the equilibrium equation \eqref{f2:A3},  takes the form of a restriction on the triplet $(\alpha,\varphi, C)$; (iv) the geometric relation \eqref{strain}, combined with \eqref{L2A}, takes the form of a restriction on the triplet $(\alpha,\varphi, \varepsilon)$, where, we recall, $\varepsilon$ is the axial strain; (v) the system of these two restrictions provides implicit representations in terms of the pair $(\varepsilon,C)$ for both $\alpha$ and $\varphi$.
All in all, we have from \eqref{forzab} that
\begin{equation}\label{forzac}
\Fc=\widehat\Fc(\varepsilon,C):=\widetilde\Fc\big(\widetilde a(\alpha,\varphi),\widetilde b(\alpha,\varphi,C),\alpha,\widetilde\beta(\alpha,\varphi),\widetilde\Theta_1(\alpha,\varphi),\widetilde\Theta_2(\alpha,\varphi)\big), \quad
\alpha=\widetilde\alpha(\varepsilon,C),\;\, \varphi=\widetilde\varphi(\varepsilon,C).
\end{equation}
For the \textit{stretching stiffness} of an armchair graphene sheet loaded as indicated in Fig. \ref{caseAA} we take 
\begin{equation}\label{stretsti}
\Yc=\widehat\Yc(\varepsilon,C):=\partial_\varepsilon\widehat\Fc(\varepsilon,C);
\end{equation}
the derivative of $\widehat\Yc$ with respect to $C$:
\begin{equation}\label{hardening}
\Hc=\widehat\Hc(\varepsilon,C):=\partial_C\widehat\Yc(\varepsilon,C)=\partial_{C\varepsilon}^{(2)}\widehat\Fc(\varepsilon,C)
\end{equation}
measures the sensitivity of the stretching stiffness to the applied couple: 
\begin{itemize}
	\item[(C2)] \emph{bending-induced hardening of the stretching stiffness occurs whenever} $\Hc>0$.
\end{itemize}	

\subsection{Pristine stretching stiffness}\label{priststretch}
By pristine stretching stiffness of an armchair graphene sheet we mean its GC stretching stiffness, for null couple loads and small strain $\varepsilon$ in direction 2.
In the flat ground configuration, $\varphi=0$ and $\Theta_1=\Theta_2=0$, and the relevant equilibrium equations are \eqref{f1:A3}, \eqref{f2:A3} and \eqref{f3:A3}, which, with the use of \eqref{f2:A3}, takes the following form:
\begin{equation}\label{new25}
\tau_\alpha+2\tau_\beta\,\beta,_\alpha-\frac{1}{2}\sigma_b\,b\cot\frac{\alpha}{2}=0.
\end{equation}
Furthermore, equation \eqref{f2:A3}, together with the constitutive assumptions \eqref{nanostress1}-\eqref{nanostress2}  and the geometric conditions \eqref{geomcompA}  and\eqref{gamma1A}$_{1,2}$, allows to define the load per unit length as a function of $b$ and $\alpha$:
\begin{equation}\label{Fgc}
\Fc_0=\frac{F}{\lambda_{1,0}}=\frac{4}{3\sqrt{3}r_0}\,\sigma_b(b,\widetilde\beta(\alpha),\alpha)= \widetilde{\Fc}(b,\alpha);
\end{equation}
where use has been made  of the fact that the GC side length in direction 1 is
$$
\lambda_{1,0}=\frac{3}{2}n_1r_0.
$$
Equations \eqref{epsi}, \eqref{f1:A3}, and \eqref{new25}, together with the constitutive assumptions \eqref{nanostress1}$_{1}$-\eqref{nanostress2}, implicitly define each of 
$a,b$, and $\alpha$, as a function of $\varepsilon$. All in all, we have from \eqref{Fgc}:
\begin{equation}
\Fc_0=\widehat{\Fc}_0(\varepsilon)=\widetilde{\Fc}_0\big(b(\varepsilon),\alpha(\varepsilon) \big);
\end{equation}
whence the following notion of GC stretching stiffness:
\begin{equation}\label{issa}
\Yc_0=\partial_\varepsilon\widehat{\Fc}_0(\varepsilon)=\partial_b\widetilde{\Fc}_0\big(b(\varepsilon),\alpha(\varepsilon) \big)\partial_\varepsilon b(\varepsilon)+\partial_\alpha\widetilde{\Fc}_0\big( b(\varepsilon),\alpha(\varepsilon) \big)\partial_\varepsilon \alpha(\varepsilon)
\end{equation}
(hereafter, although not indicated in the interest of notational lightness, evaluation at GC has to be understood).
As to the partial derivatives of $\widetilde{\Fc}_0$ in \eqref{issa}, we have:
\begin{equation}\label{DF}
\begin{aligned}
\partial_b\widetilde{\Fc}_0&= \frac{4}{3\sqrt{3}r_0}\kappa_b, \quad \kappa_b=\sigma_b,_b , \\ \partial_\alpha\widetilde{\Fc}_0&=\frac{4}{3\sqrt{3}\,r_0}\left(\mu_{b\alpha} -\frac{1}{2} \mu_{b\beta}\right),\quad \mu_{b\alpha}=\sigma_b,_\alpha, \;\,\mu_{b\beta}=\sigma_b,_\beta;
\end{aligned}
\end{equation}
as to the other partial derivatives in \eqref{issa}, on taking again into account equations \eqref{epsi}, \eqref{f1:A3}, \eqref{new25}, and \eqref{nanostress1}$_{1}$-\eqref{nanostress2}, we have via some cumbersome calculations that we prefer to omit that
\begin{equation}\label{de}
\begin{aligned}
& \partial_\varepsilon b=\frac{12\, r_0 \kappa_a \lambda_\alpha + 6\, r_0 \kappa_a \lambda_\beta - 
	\sqrt{3} \,r_0^2 \kappa_a \mu_{b\alpha} - 
	6\, r_0 \kappa_a \mu_{\alpha\beta} - 
	6\, r_0 \mu_{a\beta} \mu_{\beta a} - 
	12\, r_0 \kappa_a \mu_{\beta\alpha}}
{r_0^2 \kappa_a \kappa_b + 12\, \kappa_a \lambda_\alpha + 
	6\, \kappa_a \lambda_\beta - 
	3\sqrt{3}\, r_0 \kappa_a \mu_{b\alpha} - 
	6\, \kappa_a \mu_{\alpha\beta} - 6\, \mu_{a\beta} \mu_{\beta a} + 
	2 \sqrt{3}\, r_0 \kappa_a \mu_{\beta b} - 
	12\, \kappa_a \mu_{\beta\alpha}},\\
& \partial_\varepsilon \alpha=\frac{2r_0\kappa_a(\sqrt{3}r_0\kappa_b-6\,\mu_{b\alpha}+6\mu_{\beta b})}{r_0^2 \kappa_a \kappa_b + 12\, \kappa_a \lambda_\alpha + 
	6\, \kappa_a \lambda_\beta - 
	3\,\sqrt{3} r_0 \kappa_a \mu_{b\alpha} - 
	6\, \kappa_a \mu_{\alpha\beta} - 6 \mu_{a\beta} \mu_{\beta a} + 
	2\, \sqrt{3} r_0 \kappa_a \mu_{\beta b} - 
	12\, \kappa_a \mu_{\beta\alpha}},
\end{aligned}
\end{equation}
where we have set
\begin{equation}
\begin{aligned}
& \kappa_a=\sigma_a,_a,  \quad \mu_{a\beta}=\sigma_a,_\beta, \quad \mu_{\alpha b}=\tau_\alpha,_b, \quad \mu_{\beta b}=\tau_\beta,_b, \\
& \lambda_\alpha=\tau_\alpha,_\alpha, \quad \mu_{\alpha\beta}=\tau_\alpha,_\beta \quad \mu_{\beta\alpha}=\tau_\beta,_\alpha, \quad \lambda_\beta=\tau_\beta,_\beta.
\end{aligned}
\end{equation}
By combining \eqref{DF} and \eqref{de}, we arrive at:
\begin{equation}\label{compl}
\begin{aligned}
\Yc_0&=\frac{4}{3\sqrt{3}}\,\frac{  \kappa_a \Big(-6\,\mu_{b\alpha} (\mu_{b\alpha} - \mu_{\beta b} + \kappa_b \big(12 \lambda_\alpha + 6\, \lambda_\beta - 
	6\, (\mu_{\alpha\beta} + 2\, \mu_{\beta\alpha})\big)\Big)-6 \kappa_b \mu_{a\beta} \mu_{\beta a}}
{ \sqrt{3}\, r_0 \kappa_a (3\, \mu_{b\alpha} - 2\, \mu_{\beta b})-  r_0^2 \kappa_a \kappa_b + + 
	6\big (\mu_{a\beta} \mu_{\beta a} + \kappa_a (-2 \lambda_\alpha - \lambda_\beta + \mu_{\alpha\beta} + 
	2\, \mu_{\beta\alpha})\big)}\,.
\end{aligned}
\end{equation}
This expression shows that the stretching stiffness depends in a complicated way not only the \textit{bond-length stiffnesses} $\kappa_a$ and $\kappa_b$, but also on the \textit{bond-angle stiffnesses} $\lambda_{\alpha}$ and $\lambda_\beta$ and on the four \textit{coupling stiffnesses} denoted by the kernel letter $\mu$; not surprisingly, dihedral stiffness has no role.  Among other things, this result should be kept in mind in the perspective of putting together a proper homogenized theory. 

When, as we do in Sec. \eqref{rd}, we specialize the above results for the case of a 2nd-generation Brenner potential, we find that
$$
\Yc_0=242.924 \,{\rm nN/nm},
$$
a result in complete agreement with the literature (cf. \cite{Arroyo2004,Huang2006,Luhuang2009}, where the value 243 nN/nm is computed on adopting the 2nd-generation Brenner potential, by an approach completely different from ours). If $\Yc_0$ is divided by the nominal thickness $t=0.34$ nm usually adopted in the literature to evaluate graphene's Young modulus, the value 714.482 GPa is obtained.

\section{Zigzag graphene}\label{eqbeZ}
In this section we repeat the developments of the previous one, with the few changes made necessary by the different interplay between loads and geometry, so as to arrive at a complete formulation of the equilibrium problem for a zigzag graphene sheet.

\subsection{Loads and geometry}
\begin{figure}[h]
	\centering
	\includegraphics[scale=1]{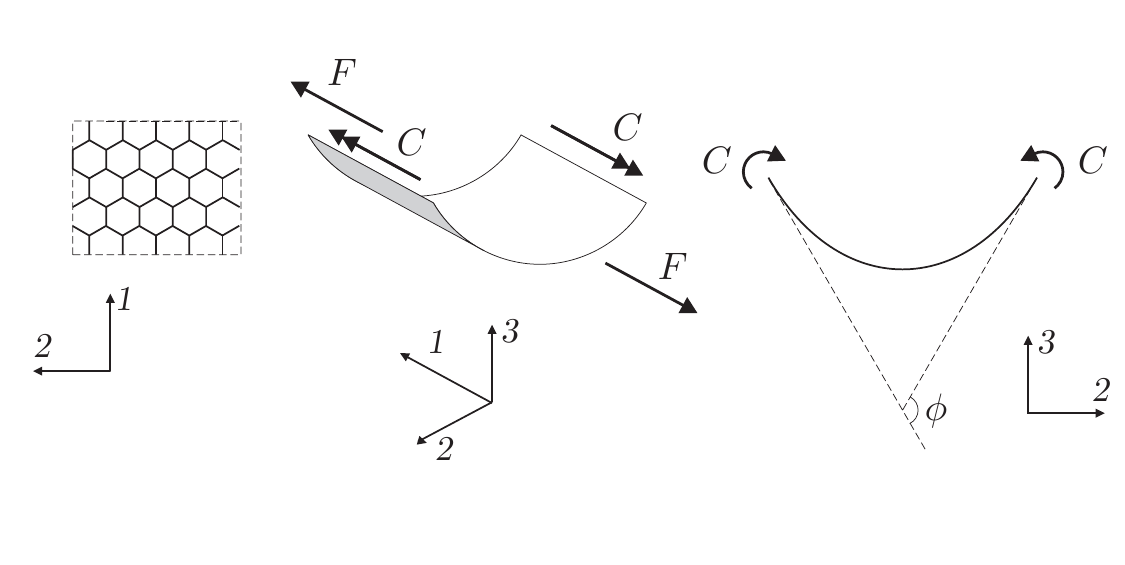}
	\caption{Bending and stretching a zigzag graphene sheet.}
	\label{caseZ}
\end{figure}

Compare  Fig.~\ref{caseZ} with Fig.~\ref{caseAA}: side couples and forces have been switched, so that now 
$C$ ($F$) is the magnitude of the resultant of each set of uniformly distributed couples (forces) acting along direction $1$ ($1$), applied to the sides parallel to direction $1$ ($2$). In view of symmetry properties that can be exploited in the same way as before, there are two inequivalent bond-lengths, two inequivalent bond-angles, and three (out of five) non-null  inequivalent dihedral angles.
The definitions of $a, b, \alpha$, and $\beta$, remain the same, and the count of the dihedral angles associated to the two types of bonds gives the same result. 	

\begin{figure}[h]
	\centering
	\includegraphics[scale=1.2]{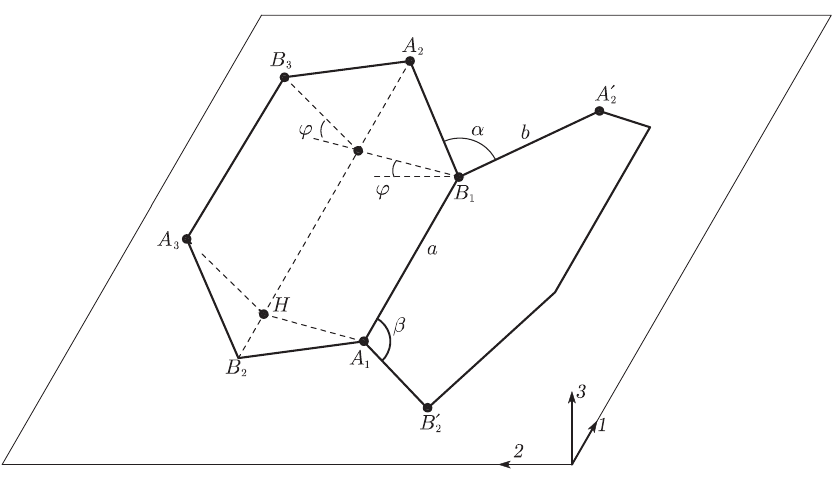}
	\caption{The deformed cell of a bended sheet in Z-direction.}
	\label{zigzag}
\end{figure}

\begin{figure}[h]
	\centering
	\includegraphics[scale=0.8]{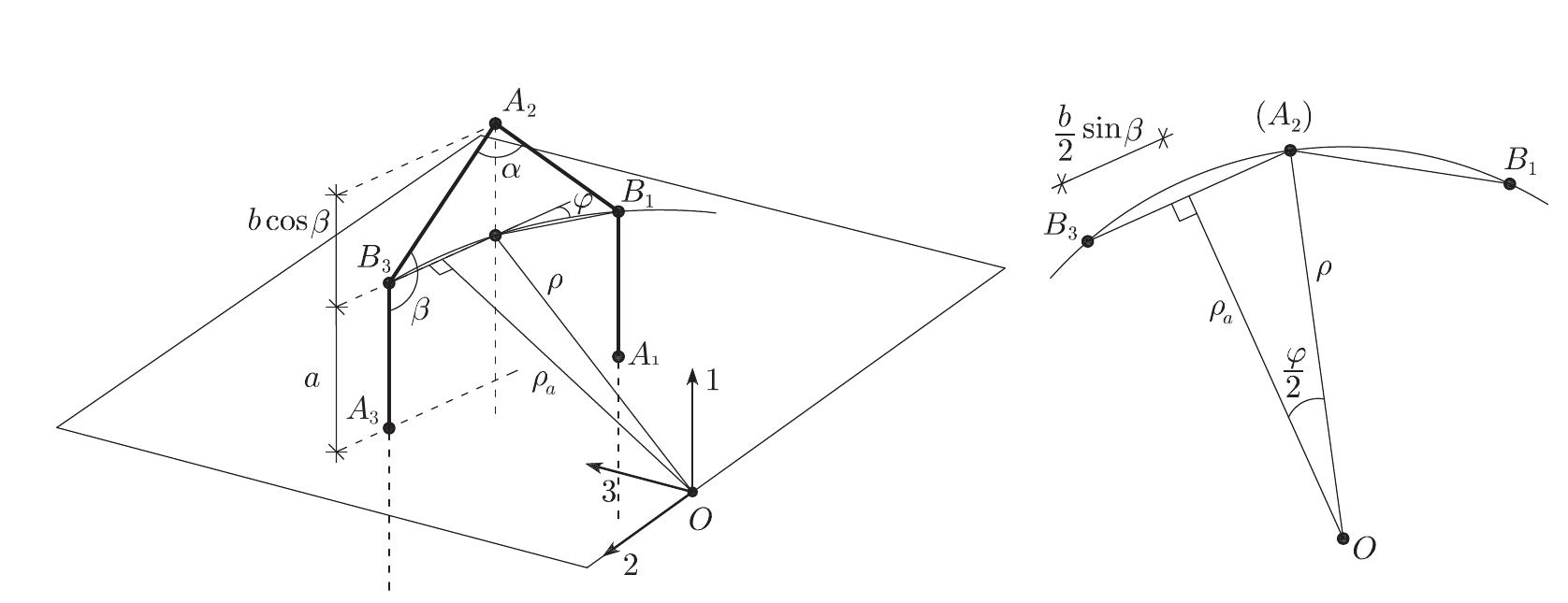}
	\caption{Local geometry of a sheet bended in Z-direction.}\label{rhoZ}
\end{figure}

The geometric conditions involving the order parameters are different from those holding in the armchair case. Let $\varphi$ denote the angle between  planes (1-2) and $A_1B_1A_2$ (see Fig. \ref{zigzag},); with reference
to the rightmost sketch in Fig.~\ref{caseZ}, we find that 
\begin{equation}\label{PhiZ}
\Phi=2n_2\varphi\,.
\end{equation}
The geometric compatibility condition for bond angles is:
\begin{equation}\label{geomcompZ}
\sin\beta\cos\frac{\varphi}{2}=\sin\frac{\alpha}{2}\,,
\end{equation}
whence
\begin{equation}\label{betafunZ}
\beta=\widetilde{\beta}(\alpha,\varphi):=\pi-\arcsin\left(\frac{\sin\displaystyle\frac{\alpha}{2}}{\cos\displaystyle\frac{\varphi}{2}}\right)\,;
\end{equation}
finally,
\begin{equation} \label{gamma1Z}
\Theta_1=\varphi\,,\quad
\sin\alpha\sin\Theta_2=\sin\beta\sin\varphi\,, \quad
\Theta_3=0\,, \quad \Theta_4=2\,\Theta_2\,, \quad \Theta_5=0\,.
\end{equation}
With the help of Fig. \ref{rhoZ}, 

\noindent it is not difficult to see that the  radius of curvature of a sheet bent in  $Z$-direction is:
\begin{equation}\label{RZ}
\rho=\frac{\sin\beta}{2\sin\varphi/2}\,b\,;
\end{equation}
the \textit{current curvature} $\kappa:=1/\rho$ has the expression
\begin{equation}\label{kappaZ}
\kappa=\left( \frac{\sin\beta}{2\sin\varphi/2}\,b \right)^{-1}.
\end{equation}
As a measure of axial deformation we take
\begin{equation}\label{strainZ}
\varepsilon=\frac{\lambda_1-\lambda_{1,0}}{\lambda_{1,0}},\quad \lambda_{1,0}=\frac{3}{2}\,n_1r_0,
\end{equation}
where $\lambda_{1,0}$ denotes the GC length of the rectangle's side in direction 1; therefore,
\begin{equation}\label{epsiZ}
\varepsilon=\frac{2}{ 3}\frac{a}{r_0}\,(1-b/a\cos\beta).
\end{equation}
\subsection{Equilibrium conditions}
By the same procedure as in Section \ref{equilarm}, the following equilibrium equations are obtained:
\begin{eqnarray}
& &\sigma_a- \frac{F}{n_2}=0\,,\label{f1:Z3}\\
& &\sigma_b + \frac{F}{2n_2}\cos\beta= 0\,,\label{f2:Z3}\\
& &\tau_\alpha+2\tau_\beta \beta,_\alpha+2\Theta_2,_\alpha(\Tc_2+\Tc_4)- \frac{1}{2}\frac{F}{n_2}b\beta,_\alpha\sin\beta= 0\,,\label{f3:Z3}\\
& & 2\tau_\beta \beta,_\varphi+\Tc_1+2\Theta_2,_\varphi(\Tc_2+ \Tc_4)-\frac{C}{n_1} = 0\,;\label{f4:Z3}
\end{eqnarray}
the constitutive equations for the nanostresses are:
\begin{eqnarray}
&&	\sigma_a  = V_R'(a)+b_a(\beta,\Theta_1)\,V_A'(a)\,,\quad
\sigma_b  = V_R'(b)+b_b(\alpha,\beta,\Theta_2,0,2\Theta_2)\,V_A'(b)\,,\label{nanostressZ1}\\
&&	\tau_\alpha  = b_b,_\alpha\!(\alpha,\beta,\Theta_2,0,2\Theta_2)\, V_A(b) \,,\quad
\tau_\beta  = \frac{1}{4}\big(b_a,_\beta\!(\beta,\Theta_1) \,V_A(a)+ 2 b_b,_\beta\!(\alpha,\beta,\Theta_2,0,2\Theta_2) \,V_A(b)\big)\,,\label{nanostressZ2}
\\
&&	\hspace{-0.8cm}\Tc_1  = \frac{1}{2}\,b_a,_{\Theta_1}\!(\beta,\Theta_1) \,V_A(a)\,,\,
\Tc_2  =  \frac{1}{2}\,b_b,_{\Theta_2}\!(\alpha,\beta,\Theta_2,0,2\Theta_2) \,V_A(b)\,,	\,\Tc_4 = \frac{1}{2}\,b_b,_{2\Theta_2}\!(\alpha,\beta,\Theta_2,0,2\Theta_2) \,V_A(b)\,.\label{nanostressZ3}
\end{eqnarray}

\subsection{Stretching-induced softening  of bending stiffness}\label{5.3}
When combined with \eqref{nanostressZ2}-\eqref{nanostressZ3} and \eqref{L2A}$_1$, equation \eqref{f4:Z3} can be used to define a mapping $\widetilde\Cc$ delivering the couple per  unit current length:
\begin{equation}\label{coppiaZ}
\Cc=\widetilde\Cc(a,b,\alpha,\beta,\Theta_1,\Theta_2):=\left(1-\frac{b}{a}\cos\beta \right)^{-1}\big(2\tau_\beta \beta,_\varphi+\Tc_1+2\Theta_2,_\varphi(\Tc_2+ \Tc_4)\big).
\end{equation}
This relation can be given a form parallel to \eqref{coppiac}, that is, the form of a function that delivers $\Cc$ as a function of $\kappa$ and $F$. 

To begin with, note that \eqref{coppiaZ} admits  a formal expression identical to \eqref{coppiab}, as the consequence of the combined implications of the following facts: 

\noindent (i) in view of the geometric relations \eqref{betafunZ} and $\eqref{gamma1Z}_{1,2}$, each of the variables $\beta$, $\Theta_1$ and $\Theta_2$  depends in a known manner on the pair $(\alpha,\varphi)$; 

\noindent (ii) in view of the equilibrium equation \eqref{f1:Z3} and of the constitutive equation  $\eqref{nanostressZ1}_1$, variable $a$ too depends in a known manner on $(\alpha,\varphi)$; 

\noindent (iii) in view of the equilibrium equation \eqref{f2:Z3} and the constitutive equation $\eqref{nanostressZ1}_2$, variable $b$ depends in a known manner on $(\alpha,\varphi)$ \emph{and} the datum $F$.

Furthermore,  on taking into account the dependences detailed just above, 

\noindent (iv) when combined with the constitutive equations $\eqref{nanostressZ2}$, the equilibrium equation \eqref{f3:Z3} takes the form of a restriction on the triplet $(\alpha,\varphi, F)$; 

\noindent (v) the geometric relation \eqref{kappaZ} takes the form of a restriction on the triplet $(\alpha,\varphi, \kappa)$, where, we recall, $\kappa$ is the current curvature; 

\noindent (vi) the system of these two restrictions provides implicit representations in terms of the pair $(\kappa,F)$ for both $\alpha$ and $\varphi$.

In conclusion, an expression of type \eqref{coppiac} is arrived at or the mapping $(\kappa,F)\mapsto\Cc$. With such a mapping at hand, we can again define as in \eqref{bendsti} the bending stiffness, and its sensitivity  to  applied tractions as in \eqref{softening}, so that  condition (C1) of Section \ref{softbend} continues to hold; the same is  true for formula \eqref{bendGS}.

\subsection{Bending-induced hardening of  stretching stiffness}\label{5.4}
When combined with \eqref{nanostressZ2}-\eqref{nanostressZ3}, equation \eqref{f2:Z3} can be used to define a mapping $\widetilde\Fc$ delivering the force per  unit current length:
\begin{equation}\label{forzaZ}
\Fc=\widetilde\Fc(a,b,\alpha,\beta,\Theta_1,\Theta_2):=-2\sigma_b\left( 2\, b\,\sin\frac{\alpha}{2} \cos\beta \right)^{-1}.
\end{equation}
Again, in view of the geometric relations \eqref{betafunZ} and $\eqref{gamma1Z}_{1,2}$, each of the variables $\beta$, $\Theta_1$ and $\Theta_2$  depends in a known manner on the pair $(\alpha,\varphi)$. Consequently,  

\noindent (i) in view of the equilibrium equation \eqref{f1:Z3} combined with \eqref{f2:Z3} and the constitutive equation  $\eqref{nanostressZ1}_1$, variable $a$  depends in a known manner on $(\alpha,\varphi)$; 

\noindent (ii) in view of the equilibrium equation \eqref{f3:Z3} and the constitutive equation $\eqref{nanostressZ1}_2$, variable $b$ depends in a known manner on $(\alpha,\varphi)$ \emph{and} the datum $C$. 

Thus, we provisionally have from \eqref{forzaZ} a formal expression identical to \eqref{forzab}. Furthermore,  in view of the dependences detailed just above, 

\noindent (iii) the equilibrium equation \eqref{f3:Z3}, when combined with the constitutive equations \eqref{nanostressZ2} and the equilibrium equation \eqref{f2:Z3},  takes the form of a restriction on the triplet $(\alpha,\varphi, C)$; 

\noindent (iv) the geometric relation \eqref{strainZ}, when combined with \eqref{L2A}$_2$, takes the form of a restriction on the triplet $(\alpha,\varphi, \varepsilon)$, where, we recall, $\varepsilon$ is the axial strain; 

\noindent (v) the system of these two restrictions provides implicit representations in terms of the pair $(\varepsilon,C)$ for both $\alpha$ and $\varphi$. 

All in all, we have an implicit representation for $\Fc$ as a function of $(\varepsilon,C)$, as in \eqref{forzac}. Therefore,  we are in a position to define the stretching stiffness as in \eqref{stretsti} and its sensitivity to applied couples as in \eqref{hardening}, as well as to state condition (C2) as in Section \ref{softyoung}; moreover,   formula \eqref{compl}  still holds true.
\vfill
\pagebreak
\section{Numerical results}\label{rd}
In this section  we collect a set of representative results and compare them with those in the literature, when available. In our computations, we choose a 2nd-generation Brenner potential, the same as in \cite{Favata2015}; the applied traction load ranges from 0 to  a value $F^{\rm max}$ approximately equal to 2/3 of graphene's fracture load ($\simeq$42 nN/nm, according to \cite{Lee2008}) and the applied couple ranges from 0 to a value $C^{\rm max}$ that induces a large curvature, approximatively equal to that of a (6-6) armchair CNT or of a (10-0) zigzag CNT.
\subsection{Stretching-induced softening of bending stiffness}\label{5.1}
\begin{figure}[h]
	\centering
	\includegraphics[scale=0.7]{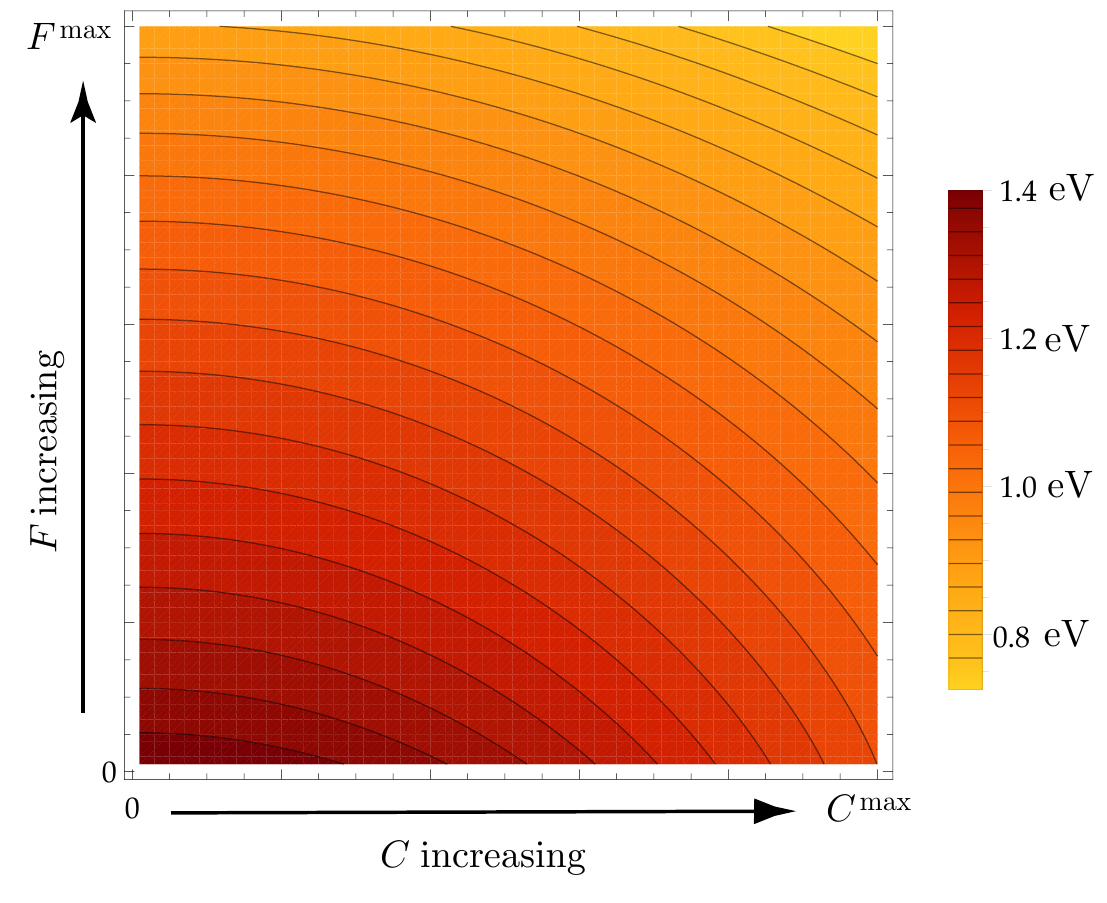}
	\quad 	\includegraphics[scale=0.7]{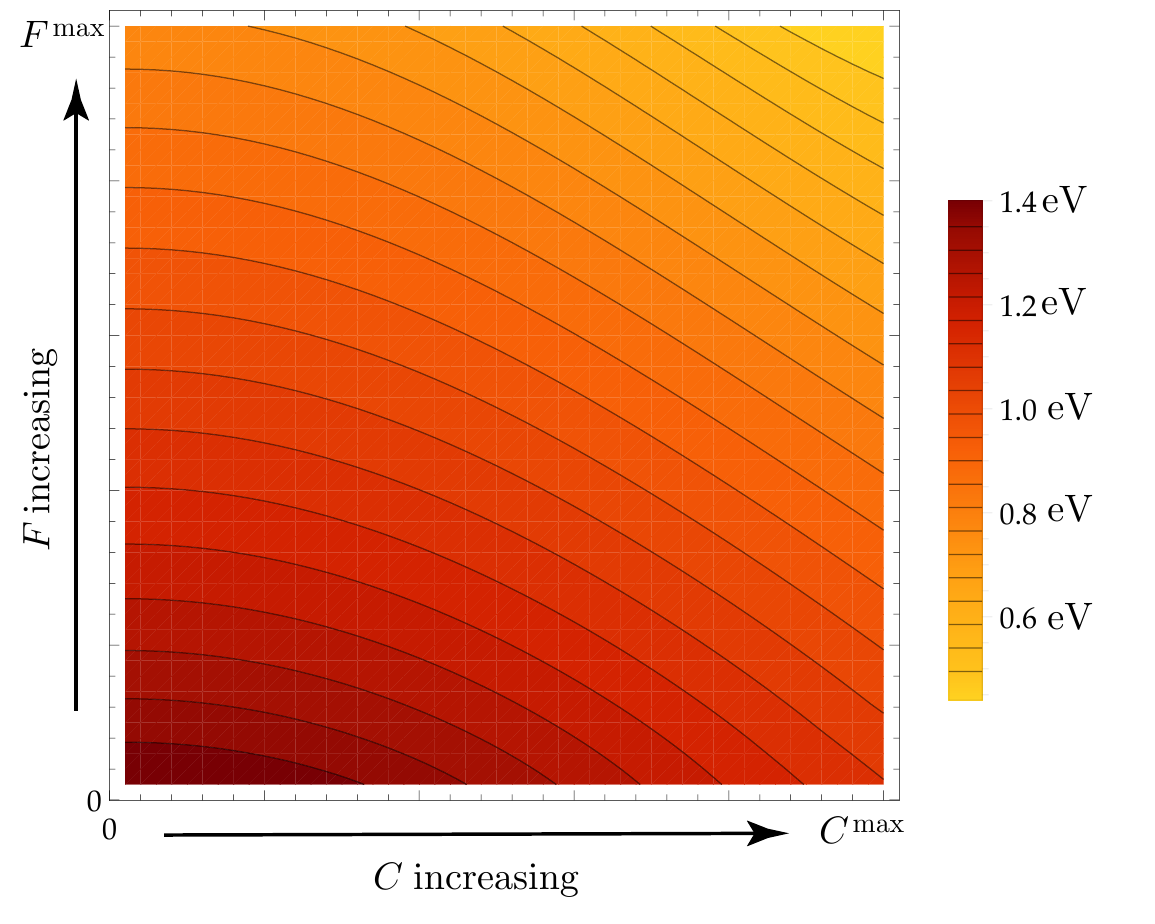}
	\caption{Level curves of bending stiffness as a function of curvature and applied traction load, armchair (left)  and zigzag (right).}
	\label{ContD}
\end{figure}

By definition (recall equation \eqref{bendsti}),  $\Dc$, the bending stiffness, is a function of  curvature $\kappa$ and applied load $F$. In the light of \eqref{kappa} (for the armchair direction) and of \eqref{kappaZ} (for the zigzag direction),  $\kappa$  can be regarded as a function of the order-parameter list $\{a,b,\alpha,\varphi\}$ that solves either system \eqref{f1:A3}-\eqref{nanostress3} (for the armchair direction) or system  \eqref{f1:Z3}-\eqref{nanostressZ3} (for the zigzag direction); and, such solution depends on the assigned data $(C,F)$. Therefore, $\Dc$ is expressible as a function of $(C,F)$, whose level curves are depicted in Fig.~\ref{ContD};
the same is true for  the axial stretch $\varepsilon$, with the use of either \eqref{epsi} (for the armchair direction) or  \eqref{epsiZ} (for the zigzag direction). 

\begin{figure}[h]
	\centering
	\includegraphics[scale=0.5]{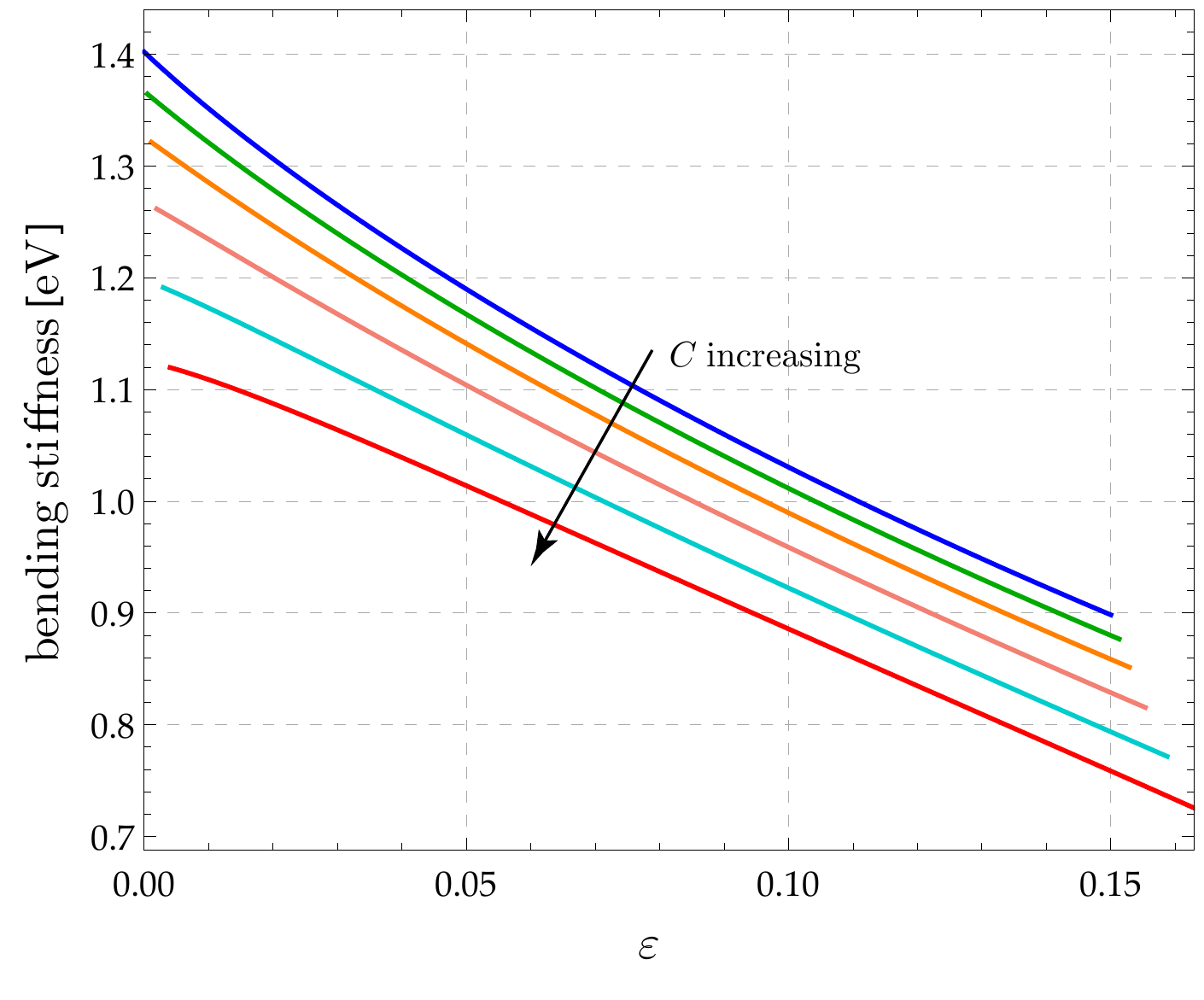}
	\quad 	\includegraphics[scale=0.5]{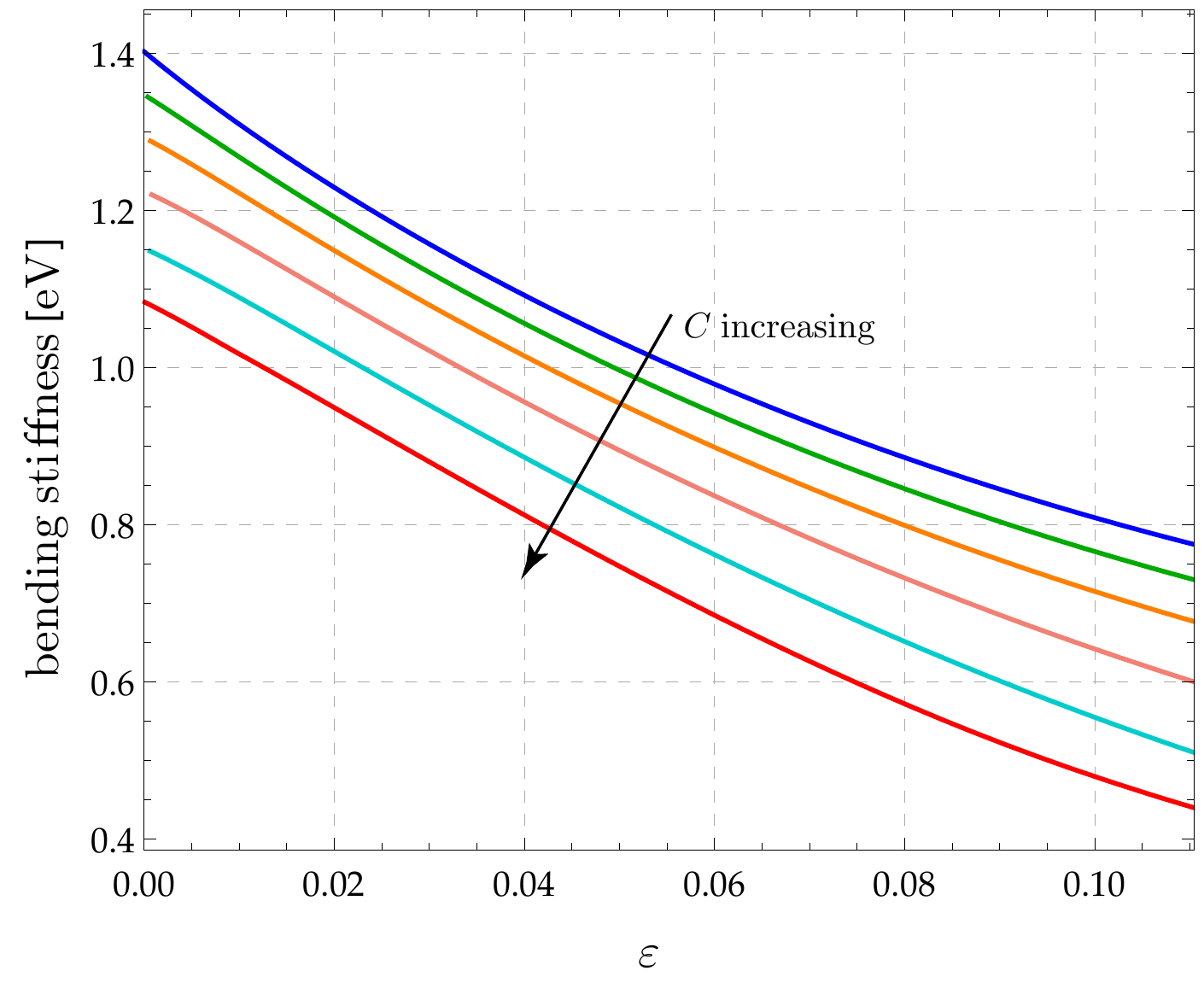}
	\caption{Bending  stiffness $\Dc$ versus axial strain $\varepsilon$, armchair (left)  and zigzag (right). }
	\label{De}
\end{figure}

Fig. \ref{De} is a plot of $\Dc=\Dc(C,F)$ vs. $\varepsilon=\varepsilon(C,F)$,
for a discrete set of values of the applied couple $C$ and for both directions, armchair  and zigzag; here and in Fig.~\ref{Se} the color code is: {{\color{blue}{\textbf{blue}}}, $C=0$; 
	{{\color{green}{\textbf{green}}}, $C=0.233$ eV;
		{{\color{orange}{\textbf{orange}}}, $C=0.346$ eV;
			{{\color{pink}{\textbf{pink}}}, $C=0.467$ eV;
				{{\color{cyan}{\textbf{cyan}}}, $C= 0.587$ eV;
					{{\color{red}{\textbf{red}}}, $C=C^{\rm max}= 0.7$ eV. Two concurrent effects can be individuated: whatever the value of parameter $C$, $\Dc$ decreases when ($F$ and hence) $\varepsilon$ increases; and, for each fixed value of $\varepsilon$, $\Dc$ decreases for increasing values of $C$. That graphene is softer to bend 
						\begin{figure}[h]
							\centering
							\includegraphics[scale=0.5]{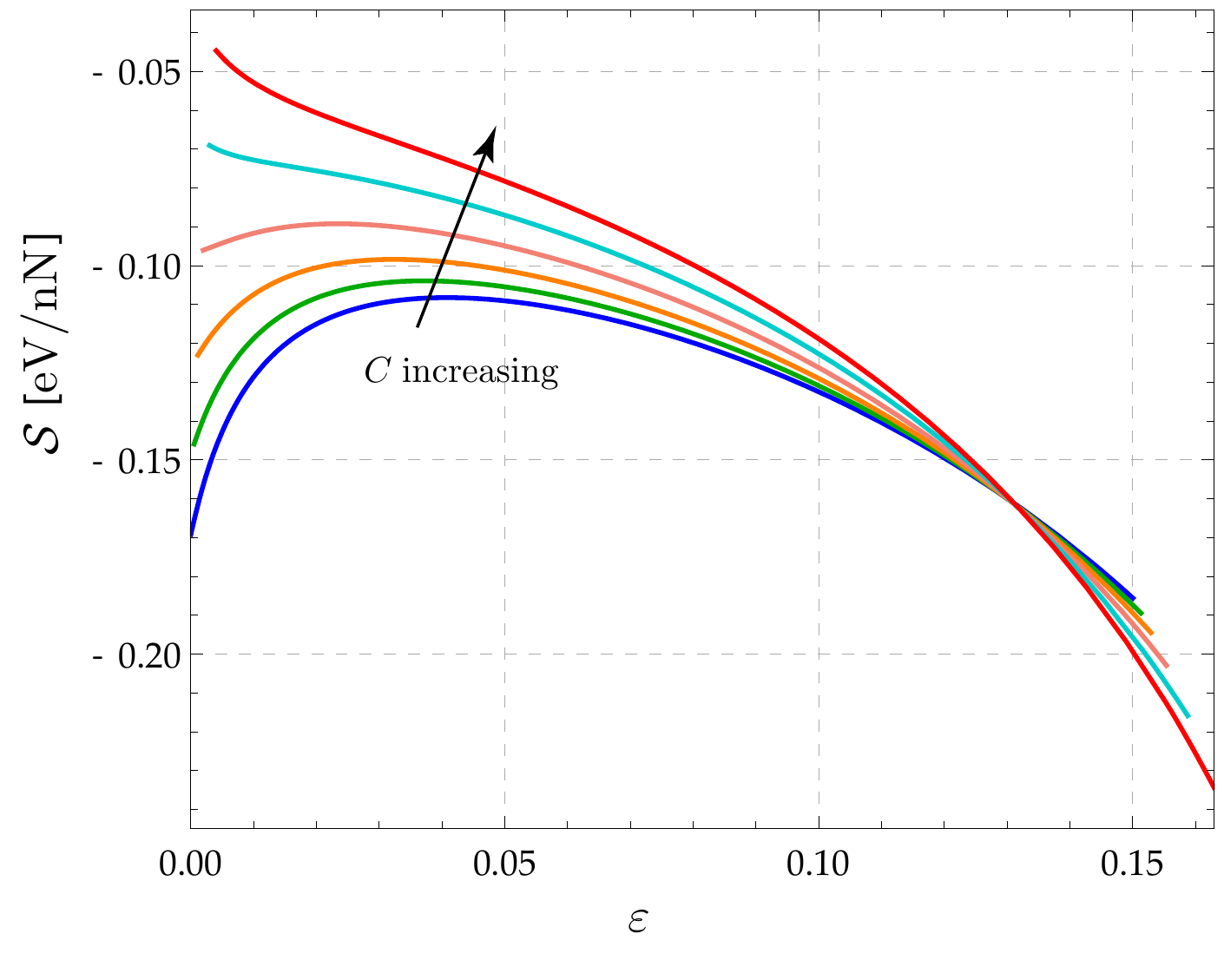}
							\quad 	\includegraphics[scale=0.5]{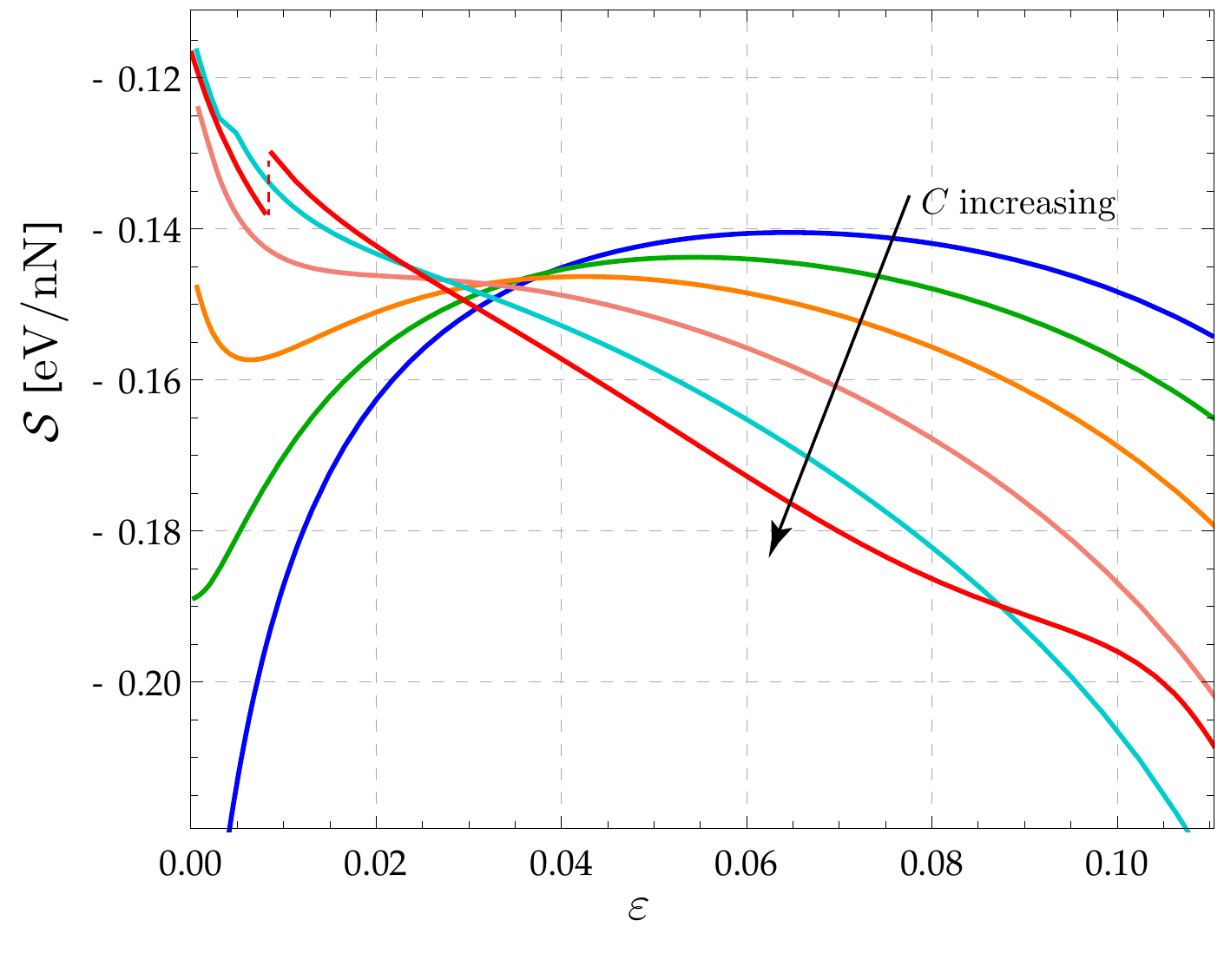}
							\caption{Softening measure $\Sc$ versus axial strain $\varepsilon$, armchair (left)  and zigzag (right).  }
							\label{Se}
						\end{figure}	
						when stretched is also illustrated by Fig. \ref{Se}, where the softening measure $\Sc$ is plotted as a function of $\varepsilon$ for the same set of values of parameter $C$
						as in Fig. \ref{De}: at first glance, we see that $\Sc$ is always negative for both directions, armchair and zigzag; we also see that, in the zigzag case,   the 2nd-generation Brenner potential predicts an unexpected jump of $\Sc$ for $C=C^{\rm max}$ ({\color{red}{\textbf{red}}} curve).
						
						As a  complement to the quantitative information about the bending stiffness summarized in Fig.~\ref{De}, we point out that
						\begin{enumerate}[(i)]
							\item for $F=F^{\rm max}$, increasing the applied couple 
							produces a decreasing up to $\approx 19$\% in the armchair case and $\approx 43$\% in the zigzag case;
							\item for $F=0$, increasing the applied couple produces a decreasing up to $\approx 20$\% in the armchair case and $\approx 23$\% in the zigzag case;
							\item for  $C=C^{\rm max}$ ({\color{red} \bf red }curve), increasing the stretching load produces a decreasing up to $\approx 35$\% in the armchair case and $\approx 59$\% in the zigzag case;
							\item  for $C=0$ ({\color{blue} \bf blue} curve), increasing the stretching load up  produces a decreasing up to $\approx 36$\% in the armchair case and up to $\approx 45$\% in the zigzag case. These results can be compared with those obtained by the use of  DFT and BOT in \cite{Shi_2012}, namely, a reduction of about 35\% (armchair) and 44\% (zigzag) of the pristine bending stiffness $\Dc_0$, whose estimated value is however quite higher than according to both our theory and MD computations, namely,  1.5292 eV instead of 1.4022 eV.
						\end{enumerate}

						In Fig. \ref{Dk}, bending stiffness is plotted vs. curvature for a discrete set of values of the stretching force; the color code is:
						{{\color{blue}{\textbf{blue}}}, $F = 0$; {{\color{lblue}{\textbf{light blue}}}, $F=4.474$ nN/nm; {{\color{green}{\textbf{green}}}, $F= 8.681$ nN/nm;
									{{\color{orange}{\textbf{orange}}}, $F= 12.895$ nN/nm; {{\color{pink}{\textbf{pink}}}, $F= 17.367$ nN/nm; {{\color{cyan}{\textbf{cyan}}}, $F=21.836$ nN/nm; {{\color{red}{\textbf{red}}}, $F= 26.050$ nN/nm.
													This figure shows  that increasing (the applied couple and hence) the curvature makes the bending stiffness decrease, with  a concurring softening effect. 

													\begin{figure}[h]
														\centering
														\includegraphics[scale=0.5]{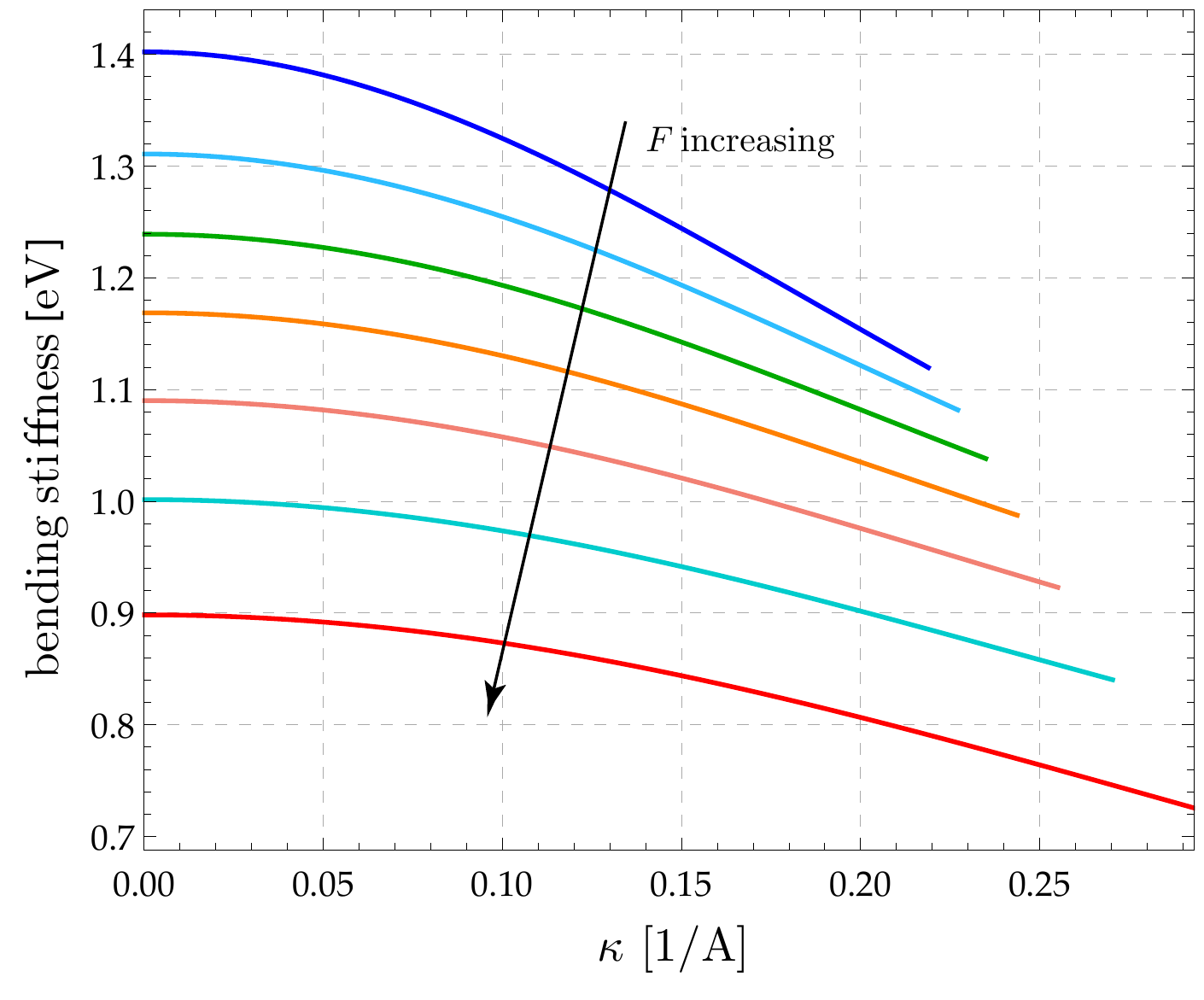}
														\quad 	\includegraphics[scale=0.5]{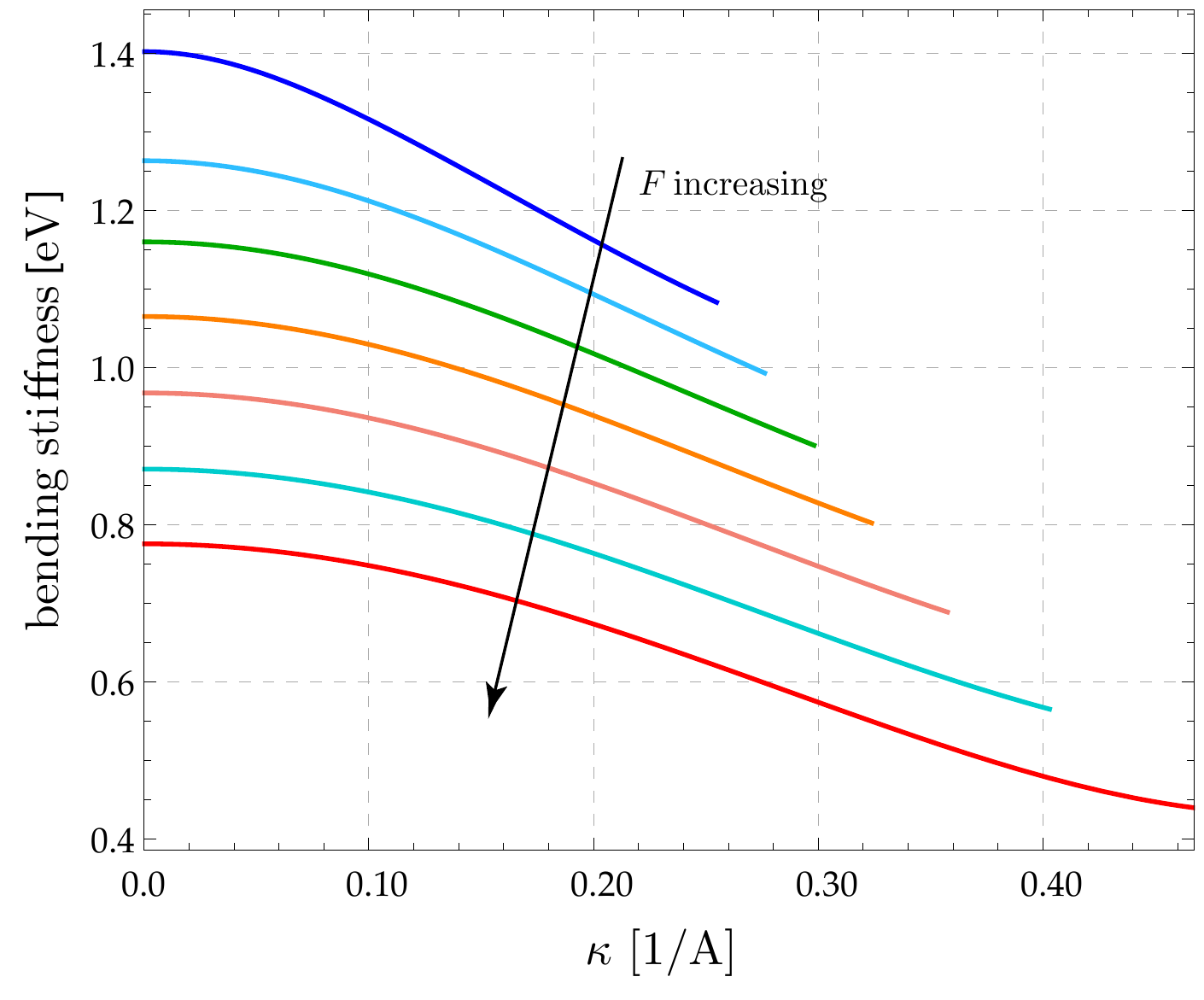}
														\caption{Bending  stiffness $\Dc$  versus curvature $\kappa$, armchair (left)  and zigzag (right).  }
														\label{Dk}
													\end{figure}

													\subsection{Bending-induced hardening of stretching stiffness}
													
													Just as detailed in the preceding subsection for the bending stiffness $\Dc$ and it sensitivity $\Sc$ to stretching, the stretching stiffness $\Yc$ and its sensitivity $\Hc$ to bending -- whose definitions are given by \eqref{stretsti} and \eqref{hardening}, respectively} -- can be expressed in terms of the assigned data $(C,F)$. The level curves of $\Yc$ as a function of $(C,F)$ are depicted in Fig.~\ref{ContY}. 
												\begin{figure}[h]
													\centering
													\includegraphics[scale=0.6]{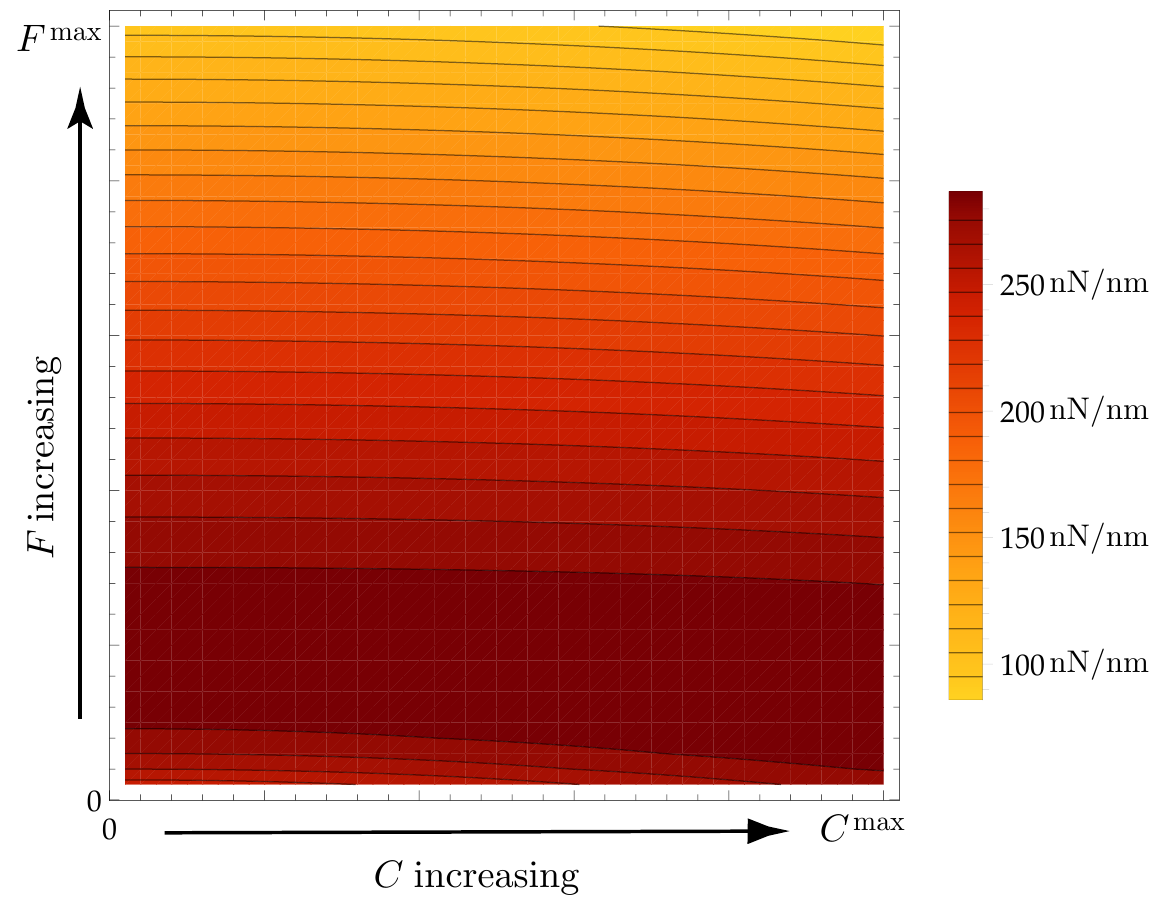}
													\quad 	\includegraphics[scale=0.6]{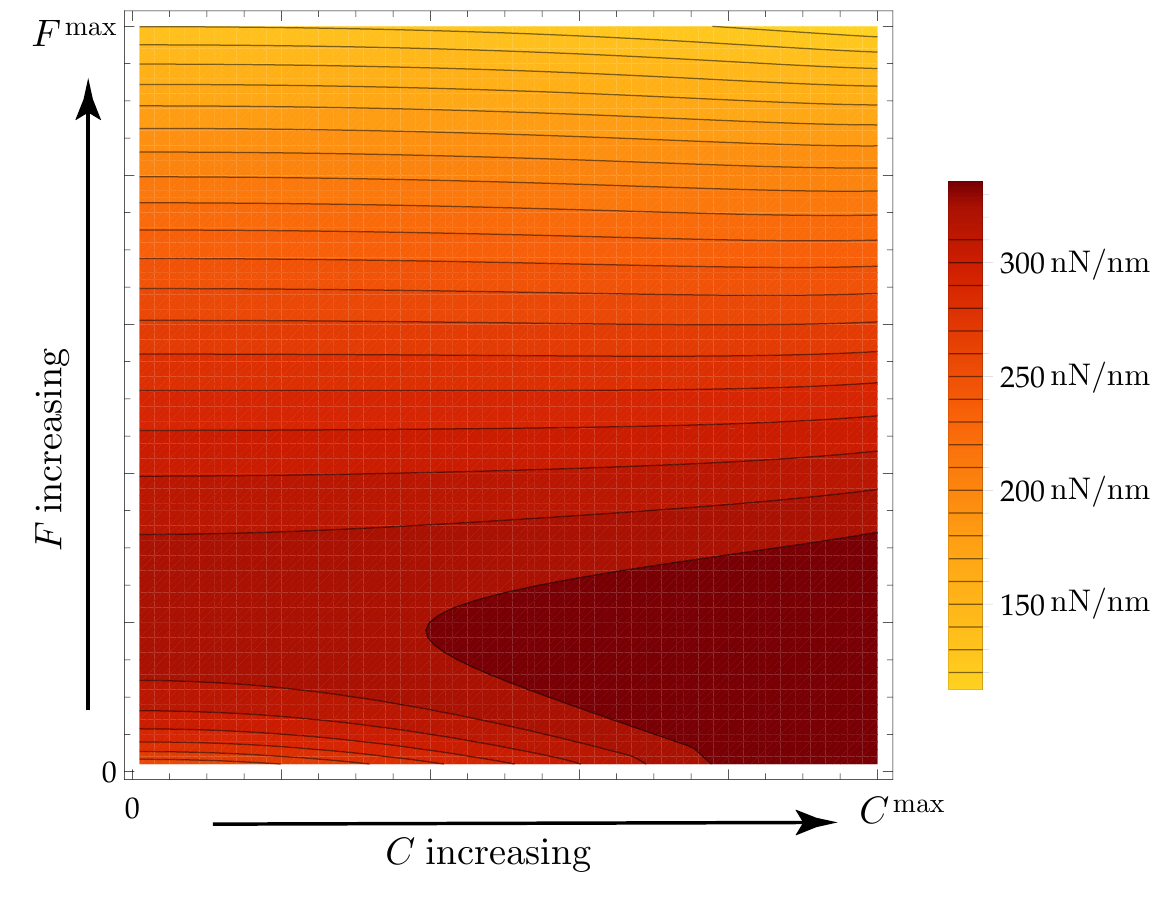}
													\caption{Level curves of stretching stiffness as a function of curvature and applied traction load, armchair (left)  and zigzag (right).}
													\label{ContY}
												\end{figure}		
												By comparison with  Fig.~\ref{ContD}, we see that the dependences of bending stiffness and stretching 
												stiffness on the data are quite different.
												Fig.~\ref{Yk} permits to visualize how the latter depends on curvature for a discrete set of increasing values of the applied traction; here and in 
												Fig.~\ref{Hk} the color code is:
												{{\color{blue}{\textbf{blue}}}, $F = 0$; {{\color{lblue}{\textbf{light blue}}}, $F=1.316$ nN/nm; {{\color{green}{\textbf{green}}}, $F= 2.632$ nN/nm;
															{{\color{orange}{\textbf{orange}}}, $F= 6.316$ nN/nm; {{\color{pink}{\textbf{pink}}}, $F= 14.472$ nN/nm; {{\color{cyan}{\textbf{cyan}}}, $F=21.836$ nN/nm; {{\color{red}{\textbf{red}}}, $F= 26.050$ nN/nm.
																			\begin{figure}[h]
																				\centering
																				\includegraphics[scale=0.5]{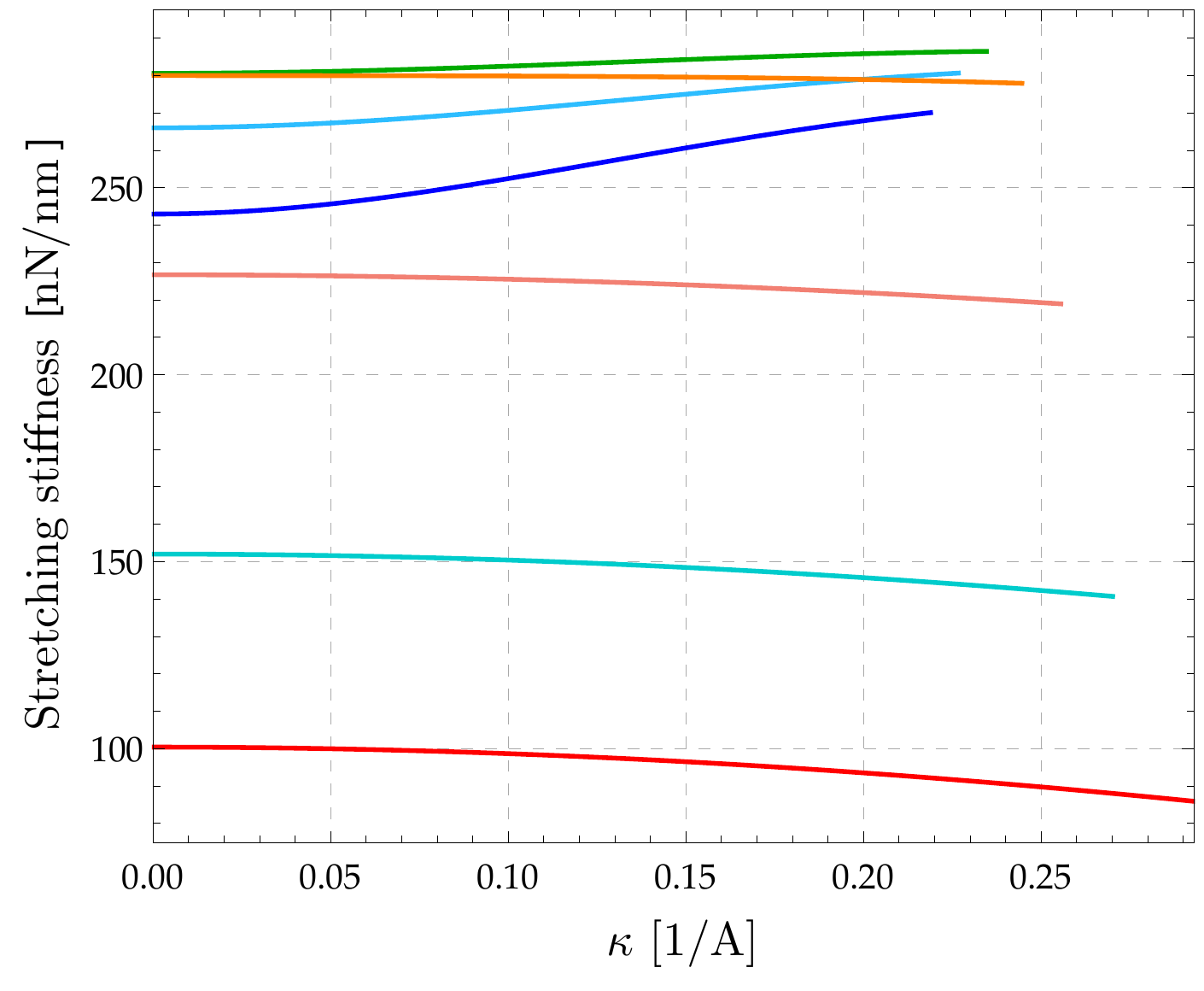}
																				\quad 	\includegraphics[scale=0.5]{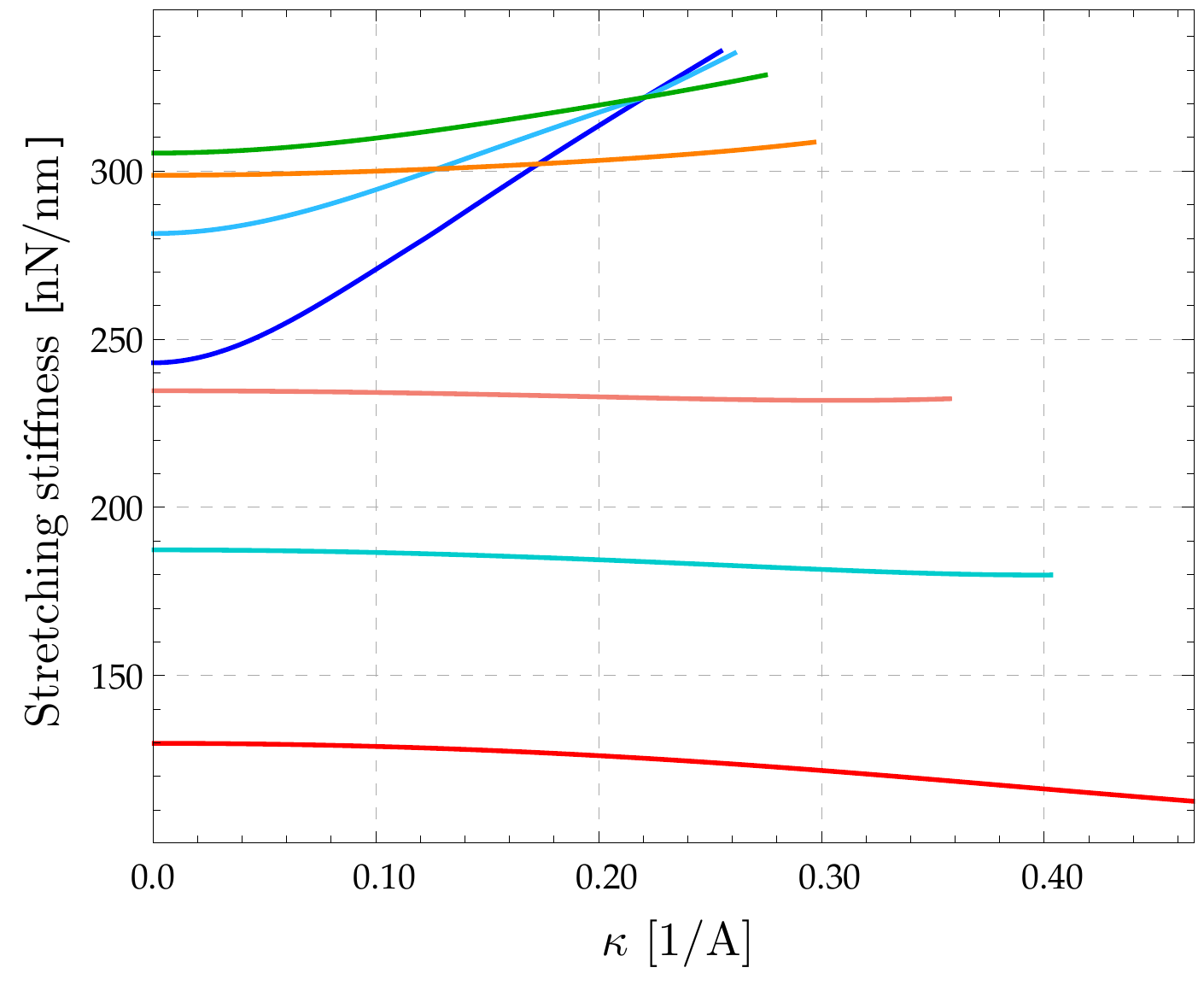}
																				\caption{Stretching stiffness $\Yc$   vs. curvature $\kappa$, armchair (left)  and zigzag (right).}
																				\label{Yk}
																			\end{figure}	
																			We see that, both for armchair and zigzag directions, bending makes graphene harder to stretch for $F=0$ ({\color{blue} \bf blue }curve) and easier to stretch for 	$F=F^{\rm max}$ ({\color{red} \bf red }curve); the transition, which is visualized by the {\color{lblue} \bf light blue}, {\color{green} \bf green } and {\color{orange} \bf orange } curves, occurs for $0<F<F^{\rm tld}$, with the threshold value $F^{\rm tld}\simeq 6.5$ nN/nm  (about  15\% of the fracture load) in the armchair case and $\simeq 12.3$ nN/nm (about  29\% of the fracture load) in the zigzag case.
																			The qualitative information summarized by Fig.~\ref{Yk} are supplemented by that in Fig.~\ref{Hk}, 
																			\begin{figure}[h]
																				\centering
																				\includegraphics[scale=0.5]{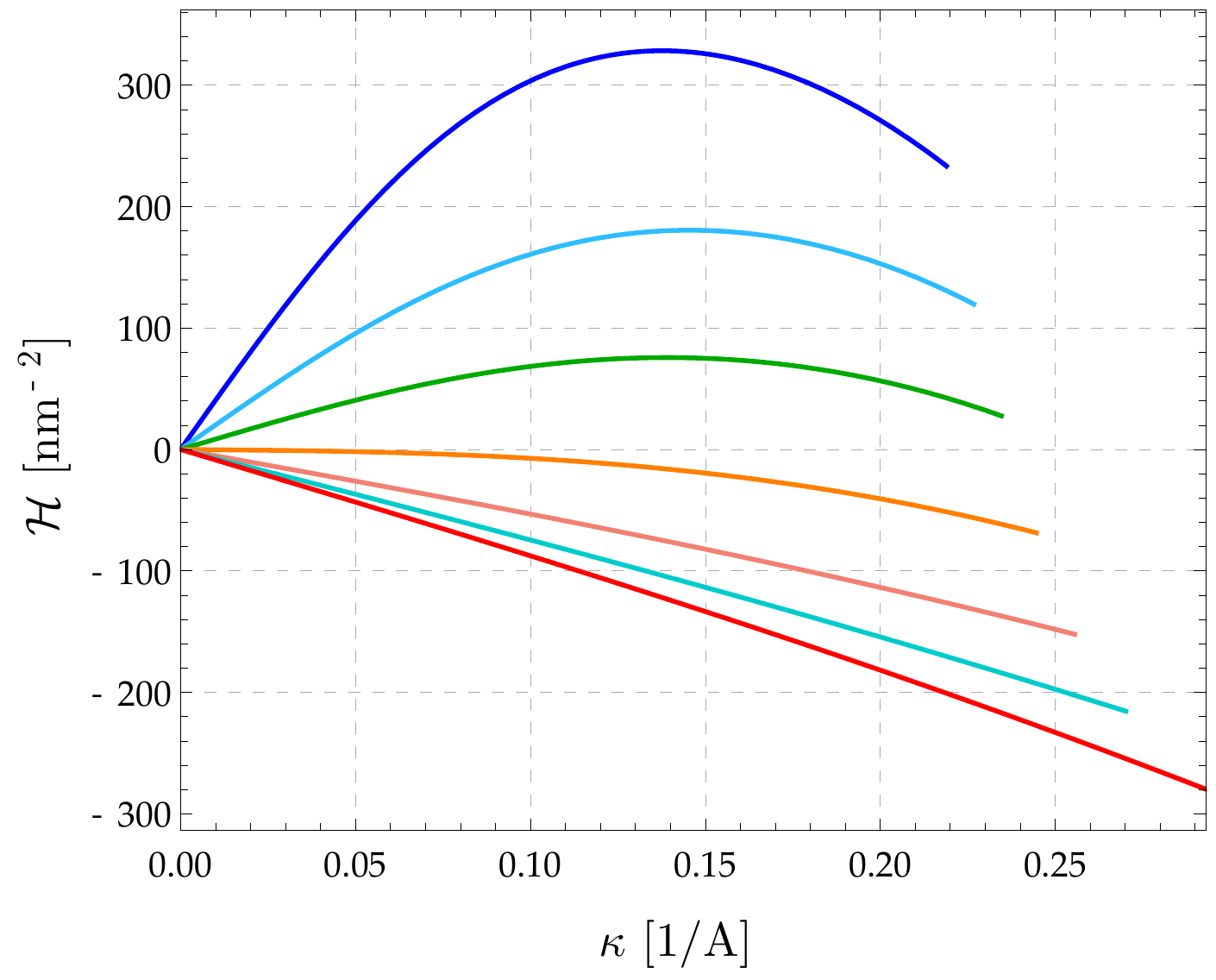}
																				\quad 	\includegraphics[scale=0.5]{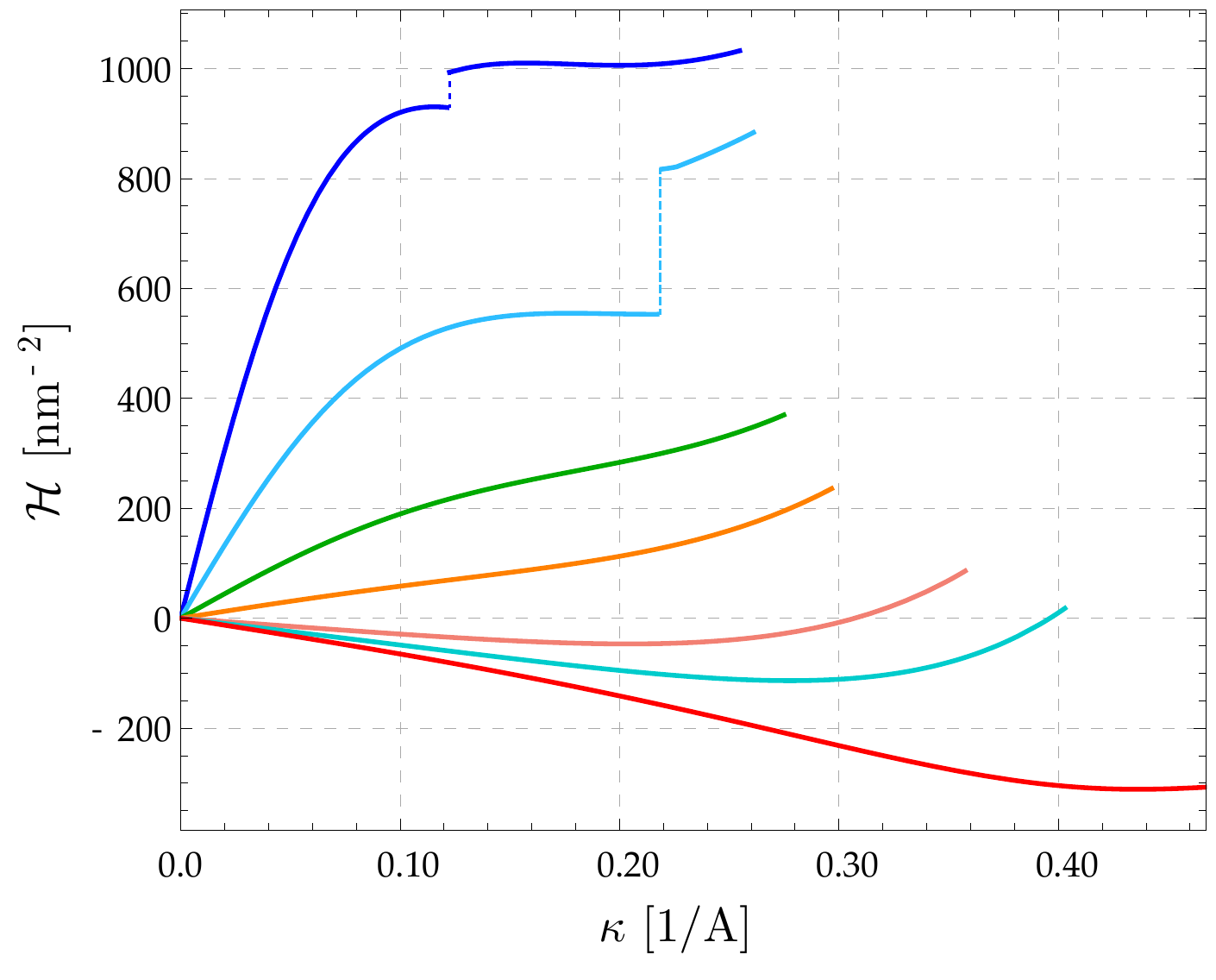}
																				\caption{Hardening measure $\Hc$ vs. curvature $\kappa$ , armchair (left)  and zigzag (right). }
																				\label{Hk}
																			\end{figure}	
																			where we also notice that once again the 2nd-generation Brenner potential predicts  unexpected jumps in the zigzag case;
																			some complementary quantitative information are:
																			\begin{enumerate}[(i)]
																				\item for $F=F^{\rm max}$ ({\color{red} \bf red }curves), the decrement in stretching stiffness is $\approx$ 14\% in the  armchair case and $\approx$ 13 \% in the zigzag case;
																				\item  for $F=0$ ({\color{blue} \bf blue }curves), the increment in stretching stiffness goes up to 
																				$\approx$ 11\% (armchair case) and $\approx$ 38\%  (zigzag case); interestingly, the stretching stiffness of bent graphene is the same as that of a CNT of identical curvature, as computed in \cite{Favata2015a}.
																			\end{enumerate}
																			\begin{figure}[h]
																				\centering
																				\includegraphics[scale=0.5]{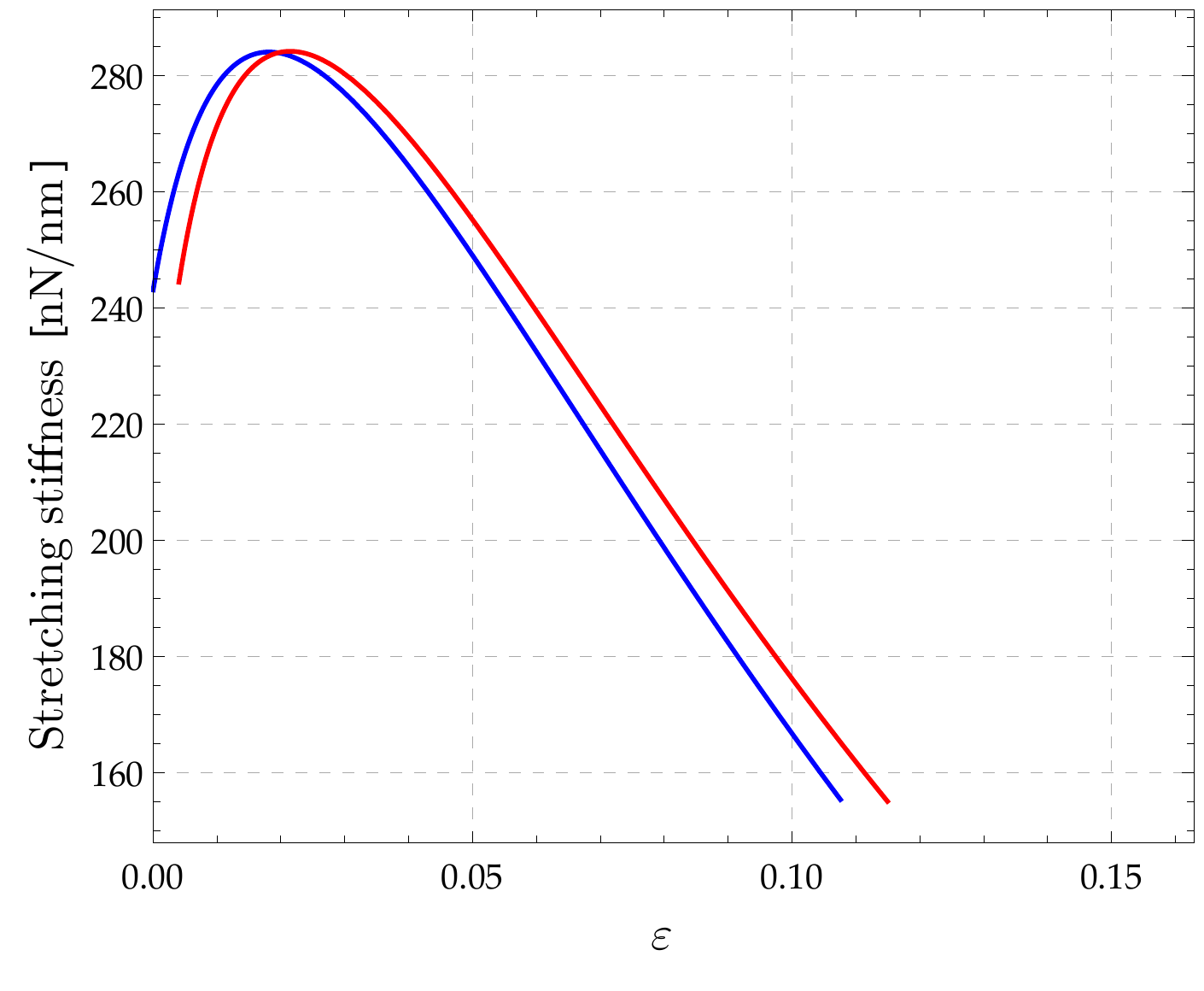}
																				\quad 	\includegraphics[scale=0.5]{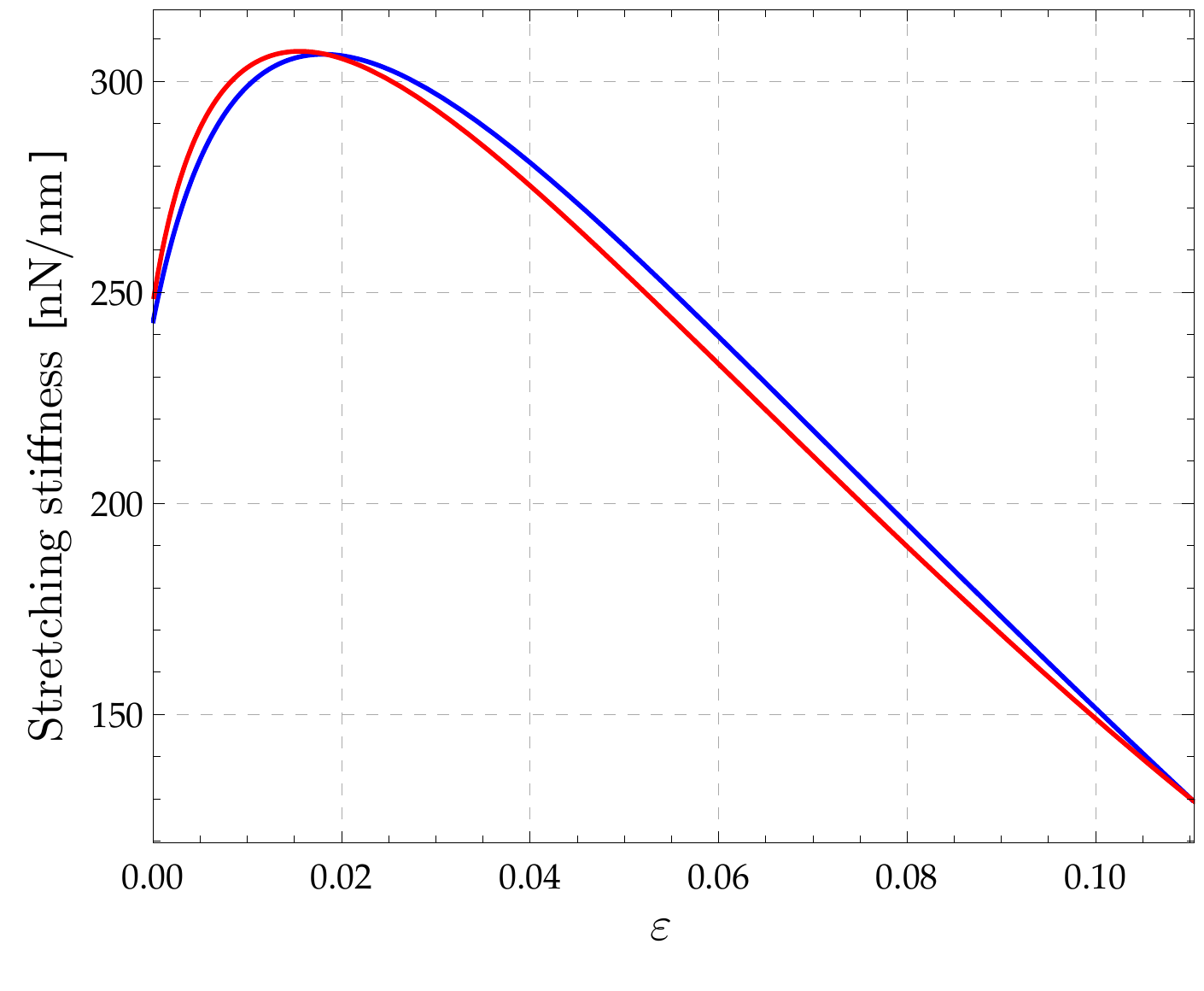}
																				\caption{Stretching stiffness  vs. $\varepsilon$, armchair (left)  and zigzag (right).  }
																				\label{Ye}
																			\end{figure}
																			While bending  makes the stretching stiffness increase, stretching makes it decrease; which of the two effects is going to prevail depends on the region of the $(C,F)$ plane one selects, in a manner that depends on the type of graphene, armchair or zigzag.  This explains why, in Fig.~\ref{Yk}, the blue and red curves do not include all the others.  This also explains why
																			in Fig.~\ref{Ye}, where the stretching stiffness is plotted vs. the amount of stretching for two fixed values of the couple, null ({\color{blue} \bf blue }curve) and large ({\color{red} \bf red }curve, $C=C^{\rm max}$), an initial hardening regime is followed by a substantial softening regime.

																			\subsection{Nanostresses}
																			Recall that we have introduced three types of nanostresses, work-conjugated to, respectively, changes in length of atomic bonds ($\sigma_a$ and $\sigma_b$), with $\sigma_a=0$ in the equilibrium problem of armchair graphene; changes in angle between two adjacent bonds ($\tau_\alpha$ and $\tau_\beta$); changes in dihedral angles ($\Tc_i$, i=1,\ldots,4), with $\Tc_4=0$ in the case of armchair graphene and $\Tc_3=0$ in the zigzag case. 
																			
																			The color code we use in the figures to follow is the same as in Subsection \ref{5.1}:  {{\color{blue}{\textbf{blue}}}, $C=0$; 
																				{{\color{green}{\textbf{green}}}, $C=0.233$ eV;
																					{{\color{orange}{\textbf{orange}}}, $C=0.346$ eV;
																						{{\color{pink}{\textbf{pink}}}, $C=0.467$ eV;
																							{{\color{cyan}{\textbf{cyan}}}, $C= 0.587$ eV;
																								{{\color{red}{\textbf{red}}}, $C=C^{\rm max}= 0.7$ eV.  We see that application of a bending couple does not 
																									\begin{figure}[h]
																										\centering
																										\includegraphics[scale=0.5]{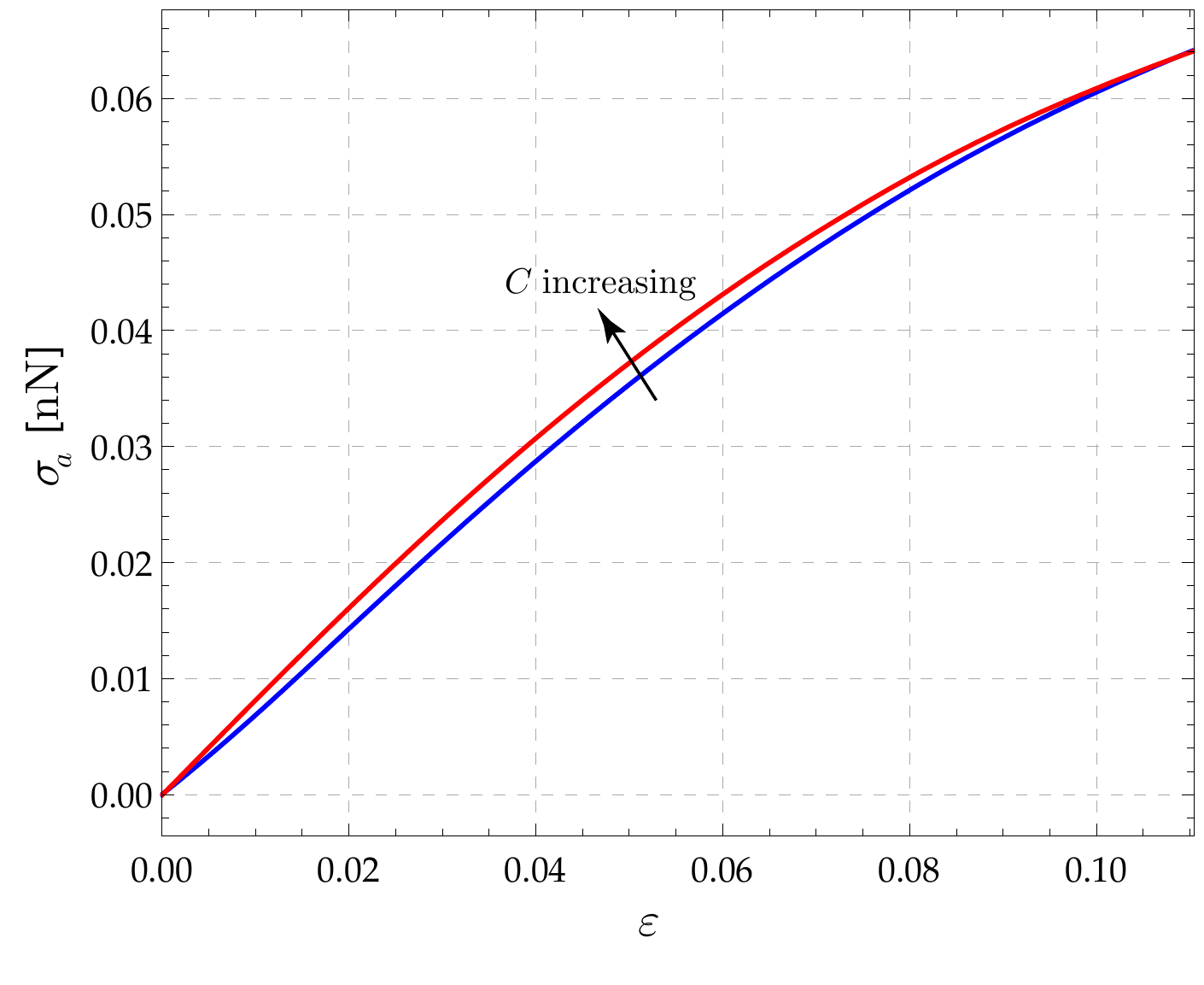}
																										\caption{Nanostress $\sigma_a$ versus  axial strain $\varepsilon$,  zigzag.   }
																										\label{sigmaa}
																									\end{figure}		
																									affect significantly either bond-length nanostresses (Fig.s \ref{sigmaa} and \ref{sigmab})
																									\begin{figure}[h]
																										\centering
																										\includegraphics[scale=0.5]{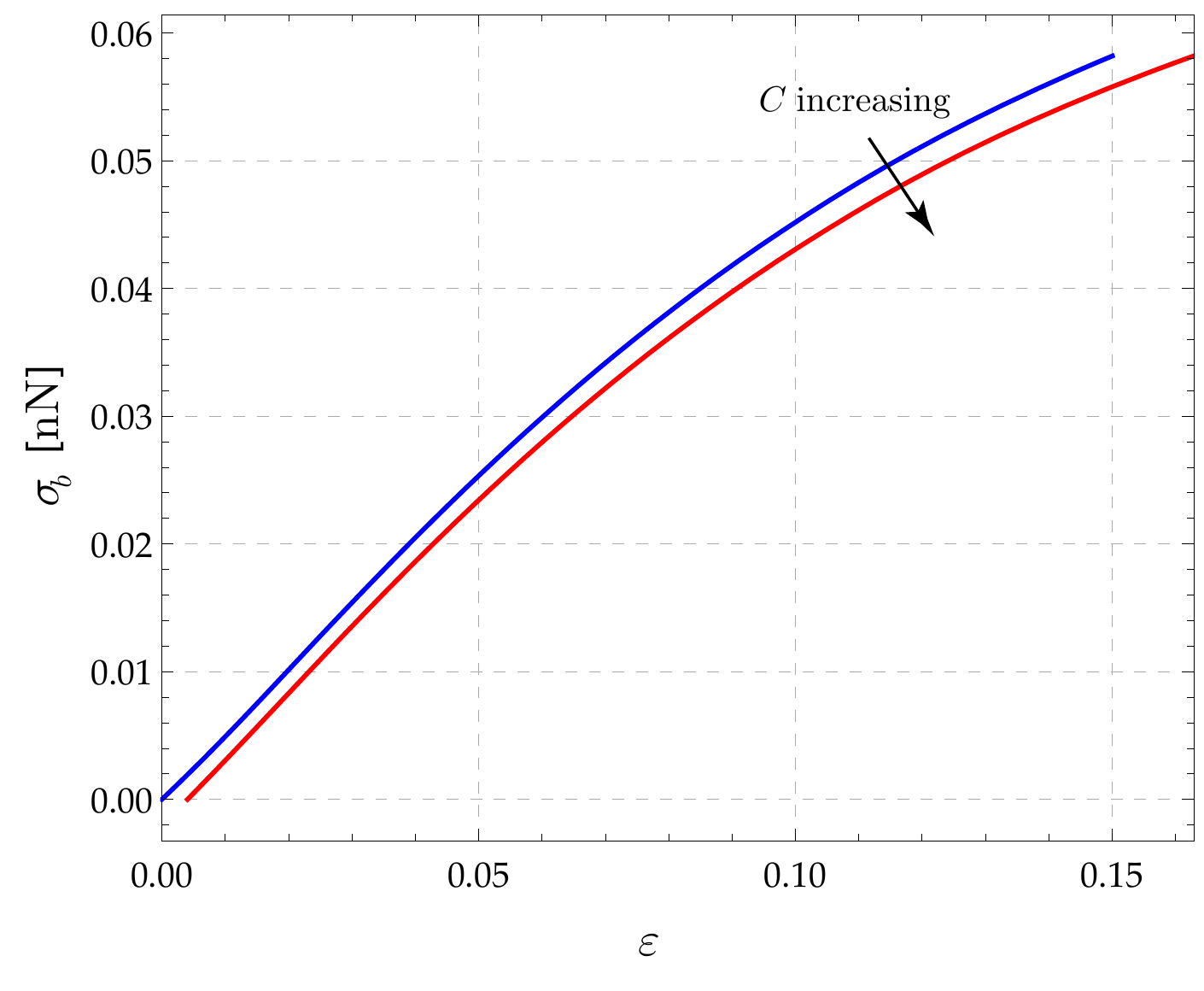}
																										\quad 	\includegraphics[scale=0.5]{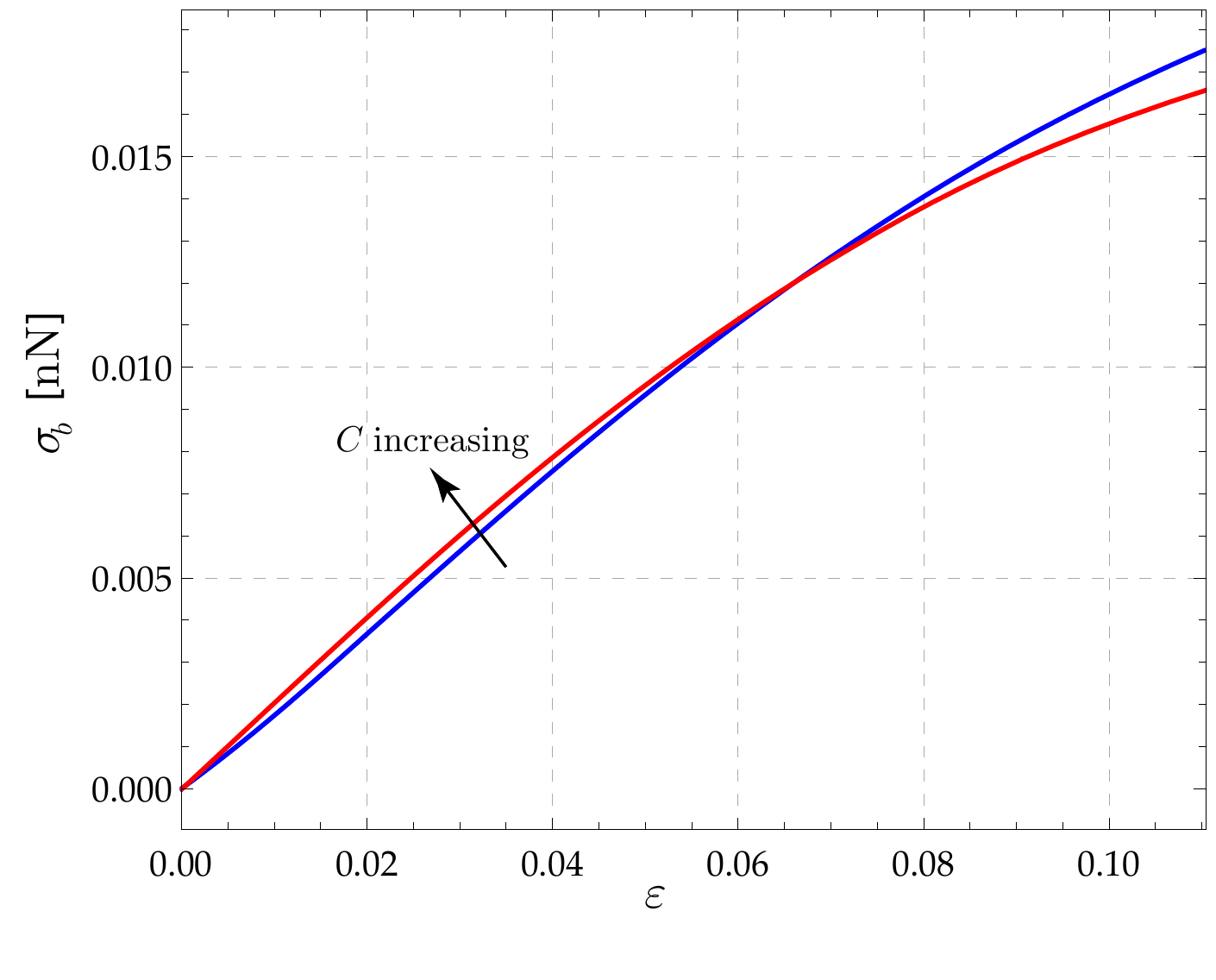}
																										\caption{Nanostress $\sigma_b$ versus  axial strain $\varepsilon$, armchair (left)  and zigzag (right; curves cross at $\varepsilon\simeq 0.65\%$). }
																										\label{sigmab}
																									\end{figure}		
																									or  the bond-angle nanostresses $\tau_a$, in the case of armchair graphene, and $\tau_b$ in the zigzag case (Fig.s \ref{taua} and \ref{taub}).
																									\begin{figure}[h]
																										\centering
																										\includegraphics[scale=0.5]{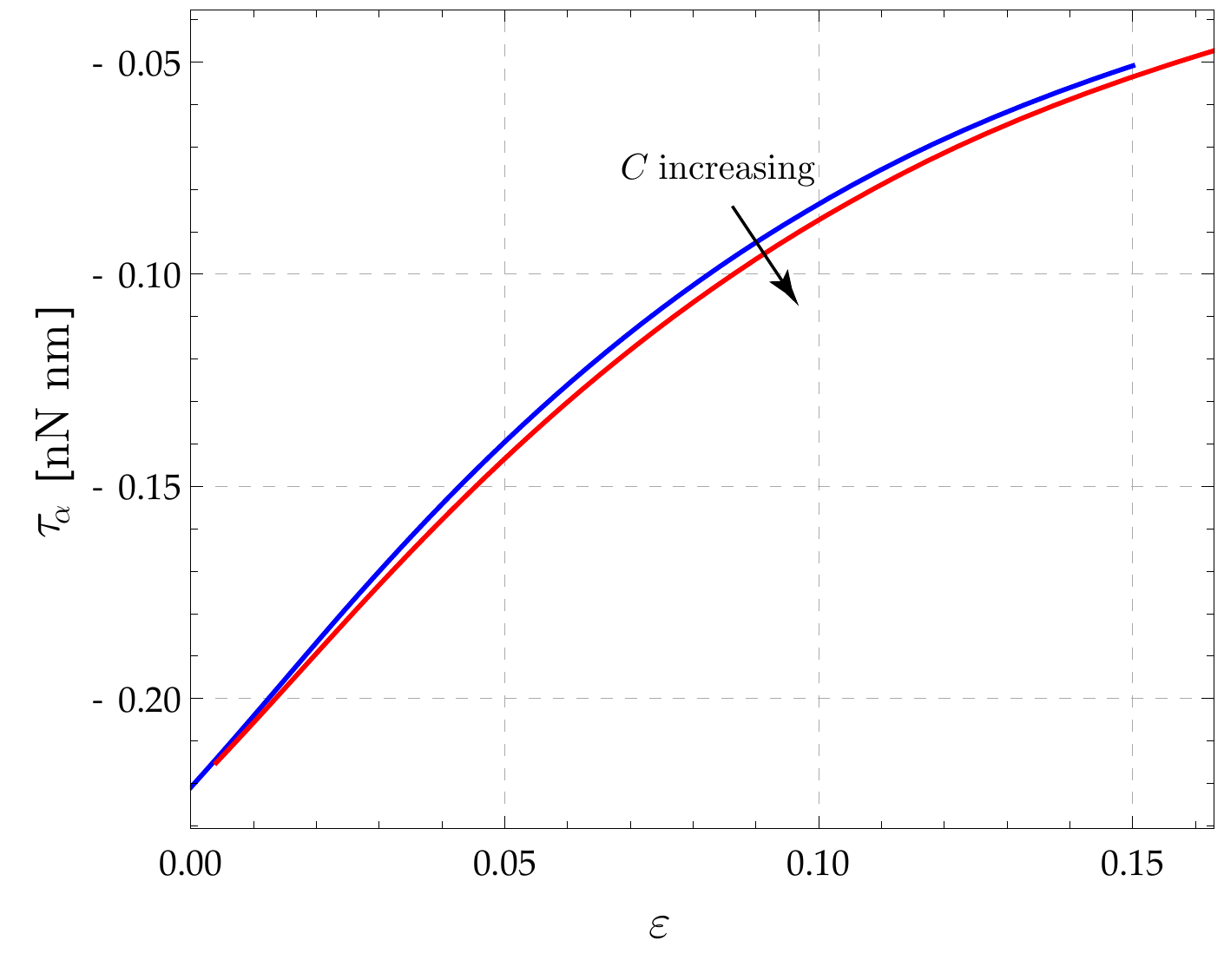}
																										\quad 	\includegraphics[scale=0.5]{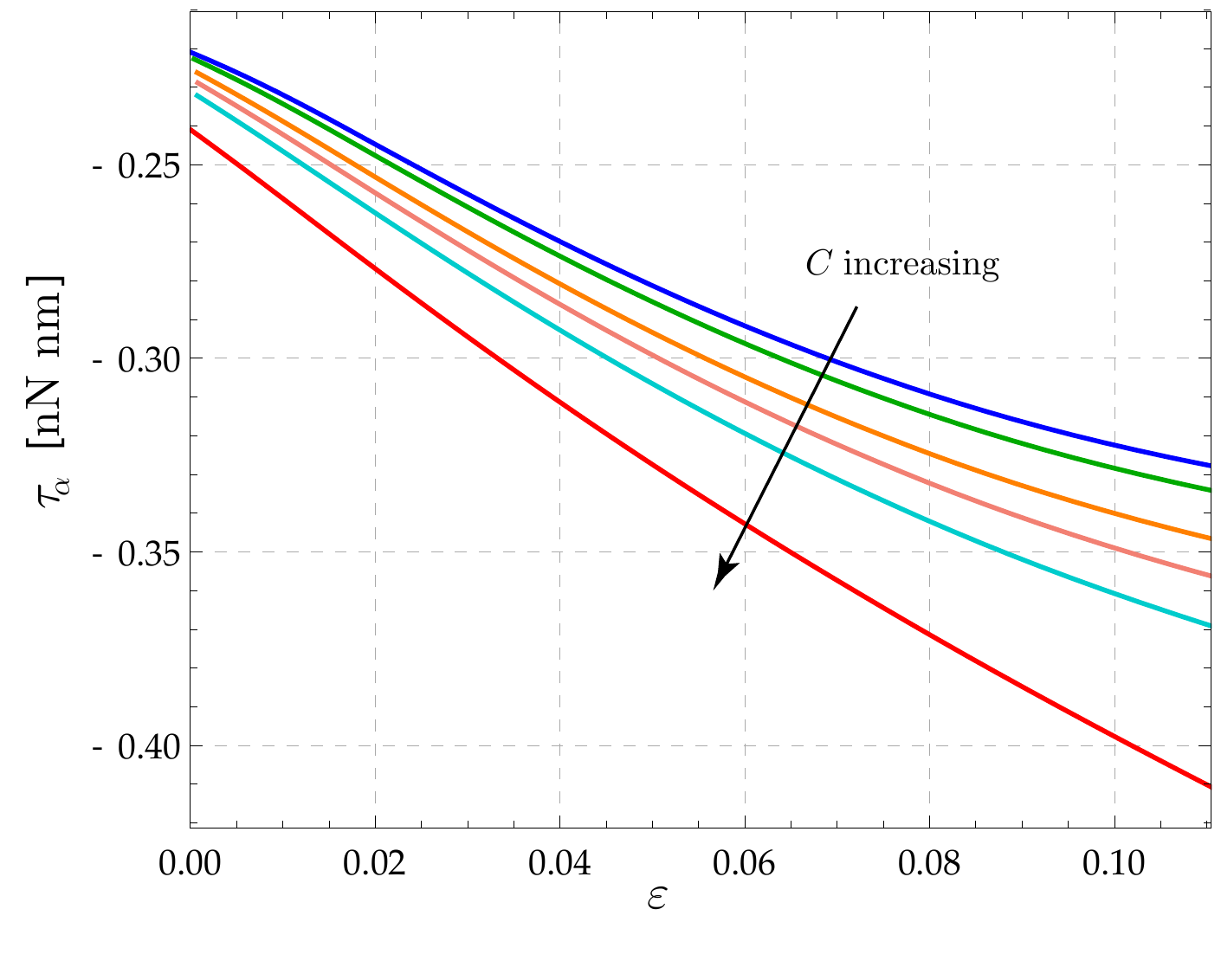}
																										\caption{Nanostress $\tau_\alpha$ versus  axial strain $\varepsilon$, armchair (left)  and zigzag (right).   }
																										\label{taua}
																									\end{figure}		
																									\begin{figure}[h]
																										\centering
																										\includegraphics[scale=0.5]{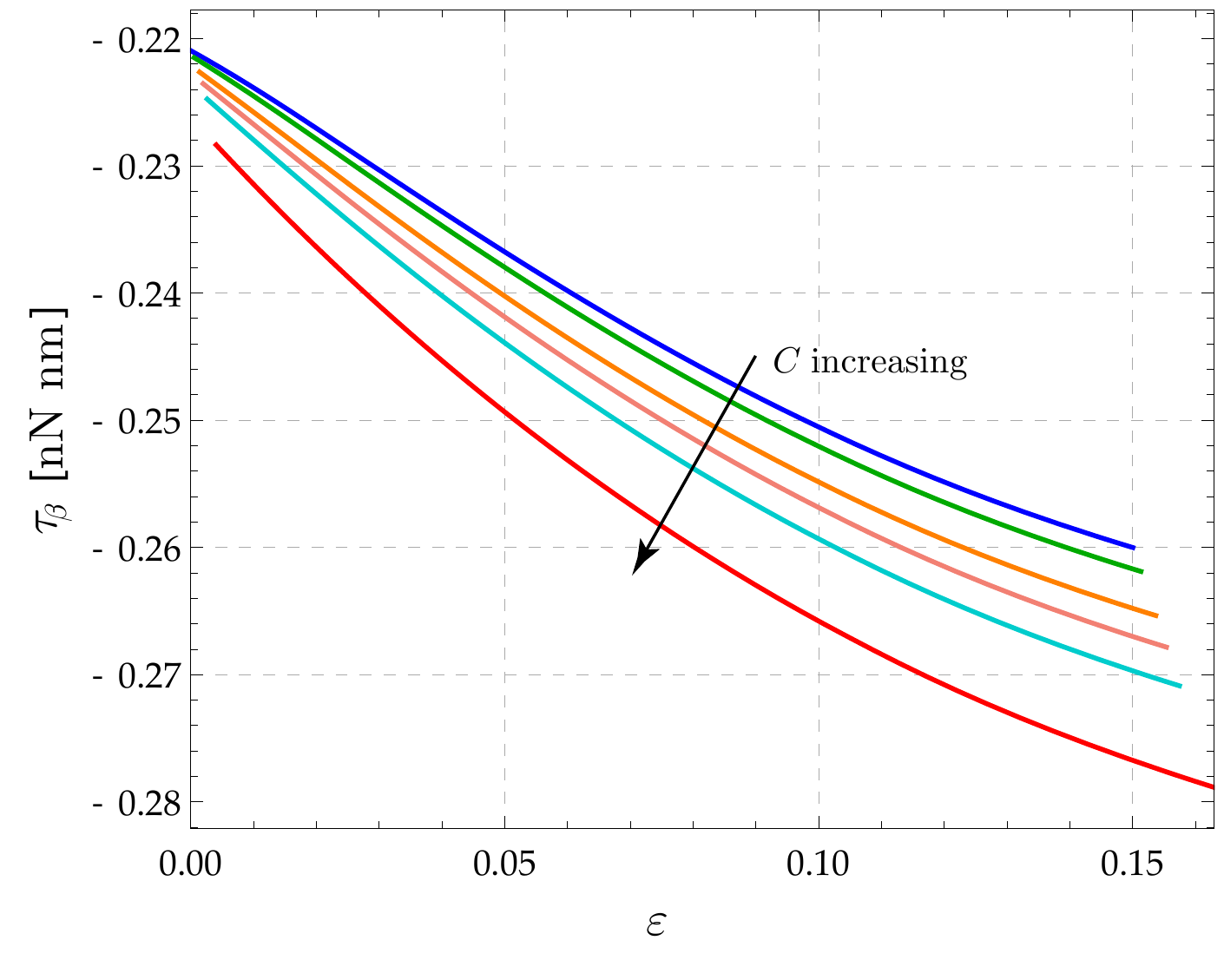}
																										\quad 	\includegraphics[scale=0.5]{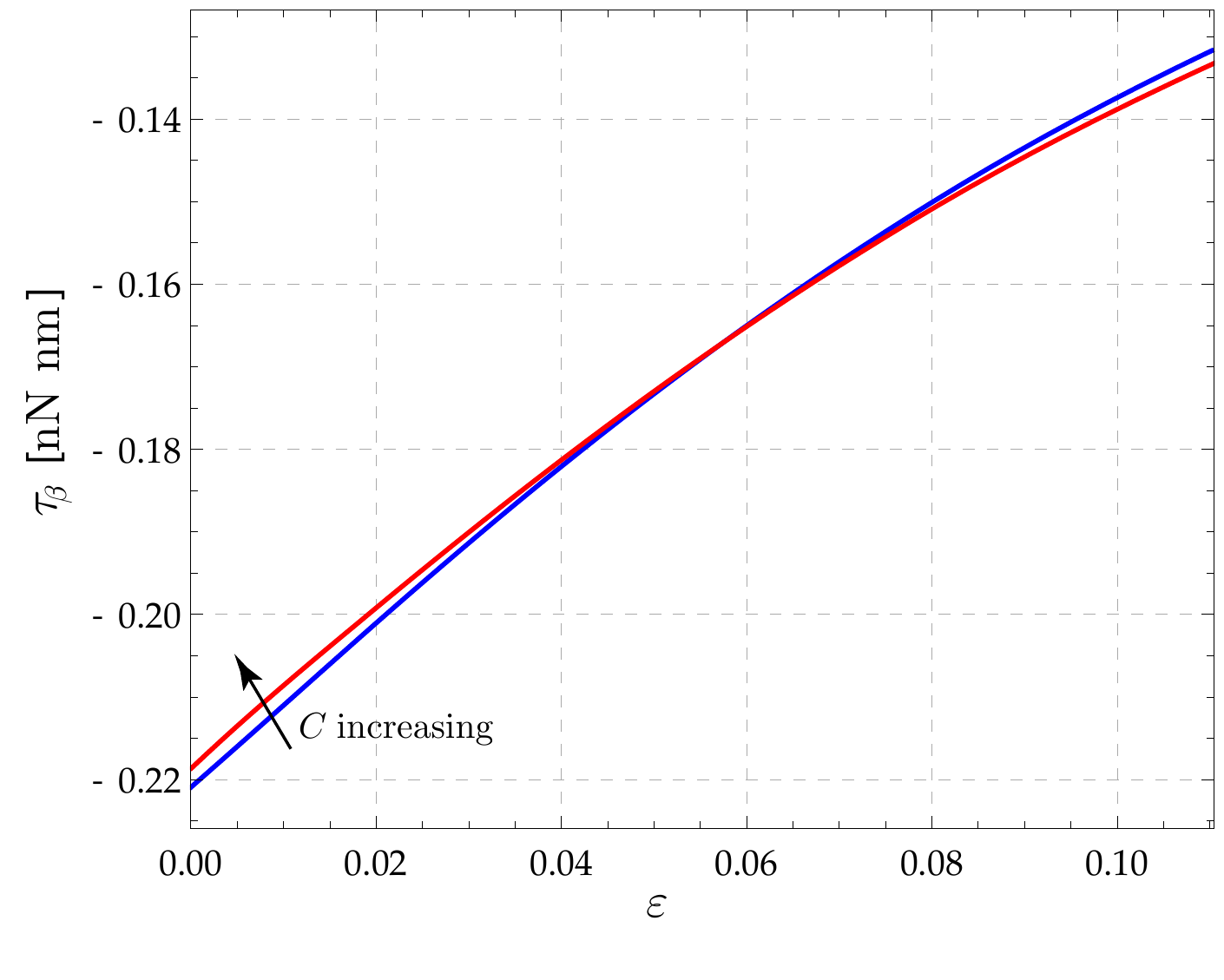}
																										\caption{Nanostress $\tau_\beta$ versus  axial strain $\varepsilon$, armchair (left)  and zigzag (right; curves cross at $\varepsilon\simeq 0.55\%$).   }
																										\label{taub}
																									\end{figure}
																									Both $\tau_\alpha$ and $\tau_\beta$ are different from zero in GC, a fact that reveals that graphene suffers a bond-angle selfstress in its ground configuration,  as extensively discussed in \cite{Favata2015}. Finally, Fig.s \ref{T1}, \ref{T2} and \ref{T3}
																									\begin{figure}[h]
																										\centering
																										\includegraphics[scale=0.5]{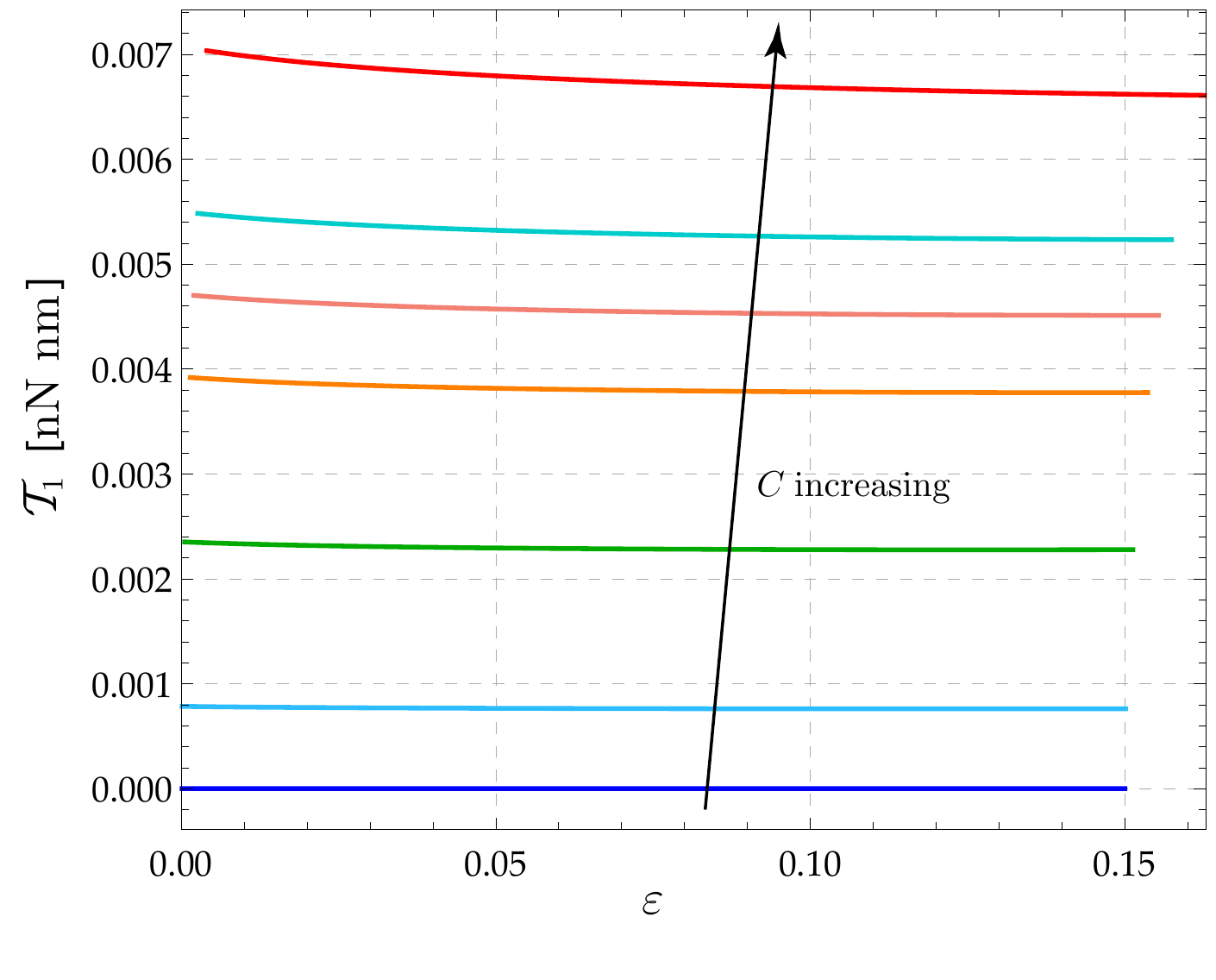}
																										\quad 	\includegraphics[scale=0.5]{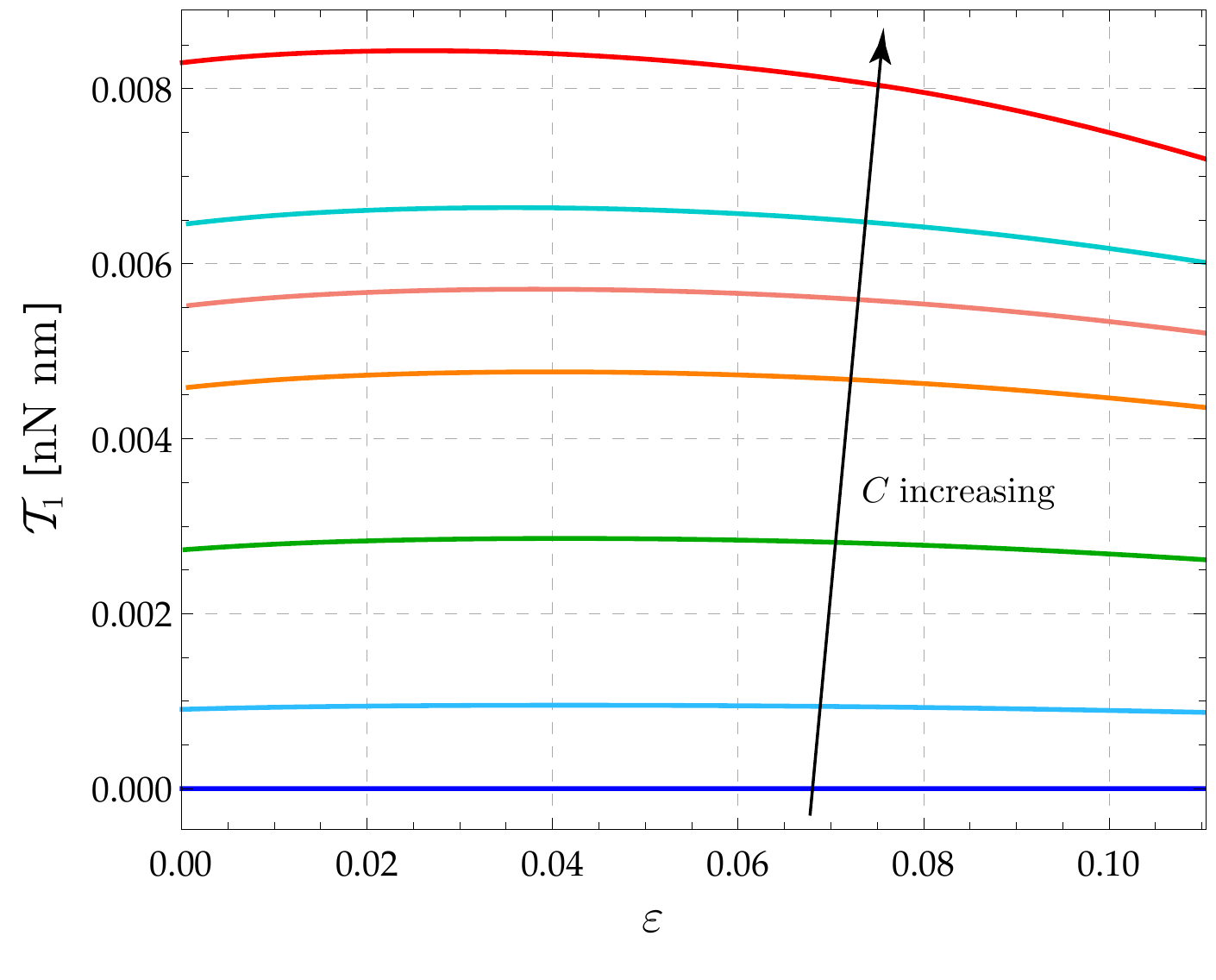}
																										\caption{Nanostress $\Tc_1$ versus  axial strain $\varepsilon$, armchair (left)  and zigzag (right).   }
																										\label{T1}
																									\end{figure}
																									\begin{figure}[h]
																										\centering
																										\includegraphics[scale=0.5]{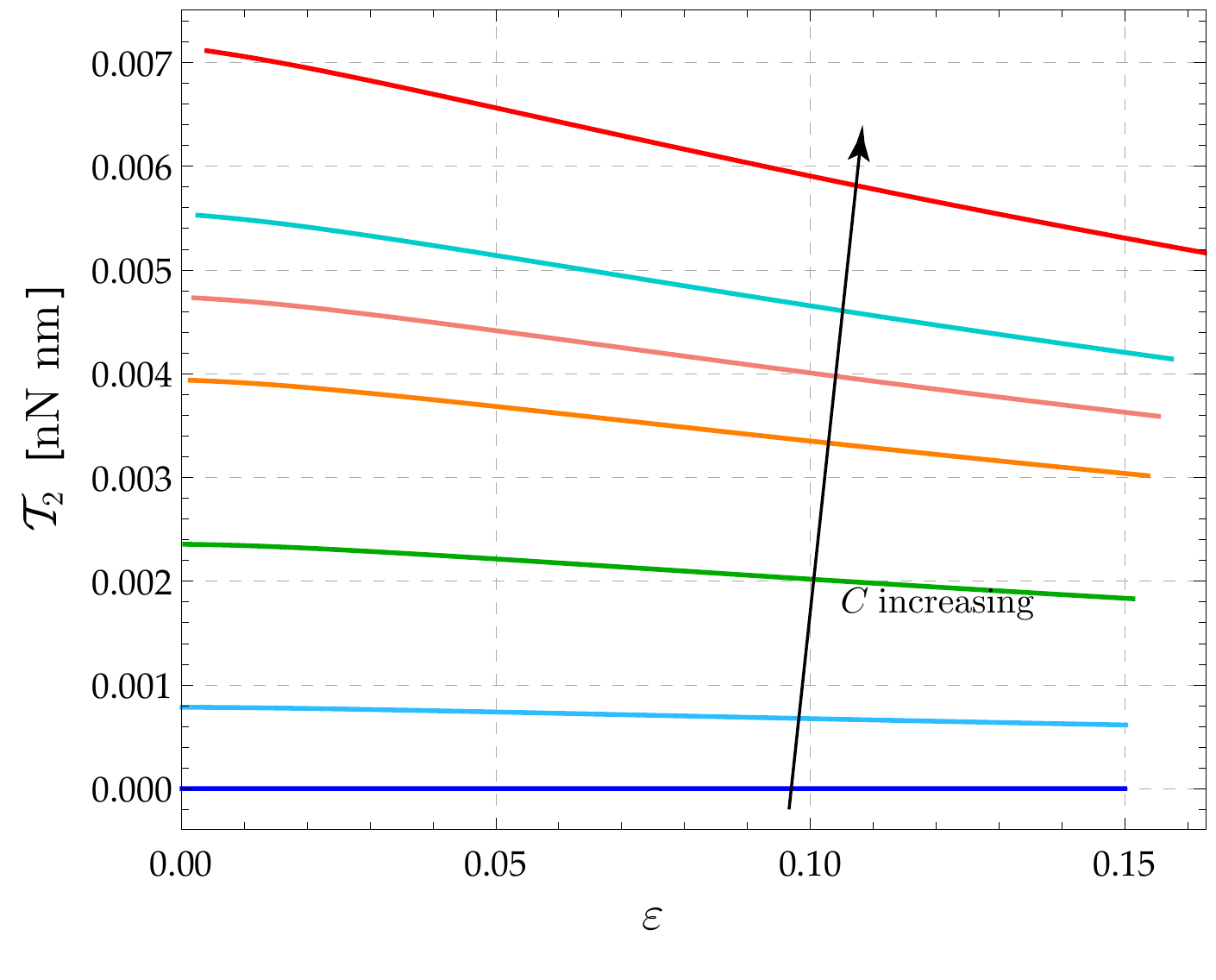}
																										\quad 	\includegraphics[scale=0.5]{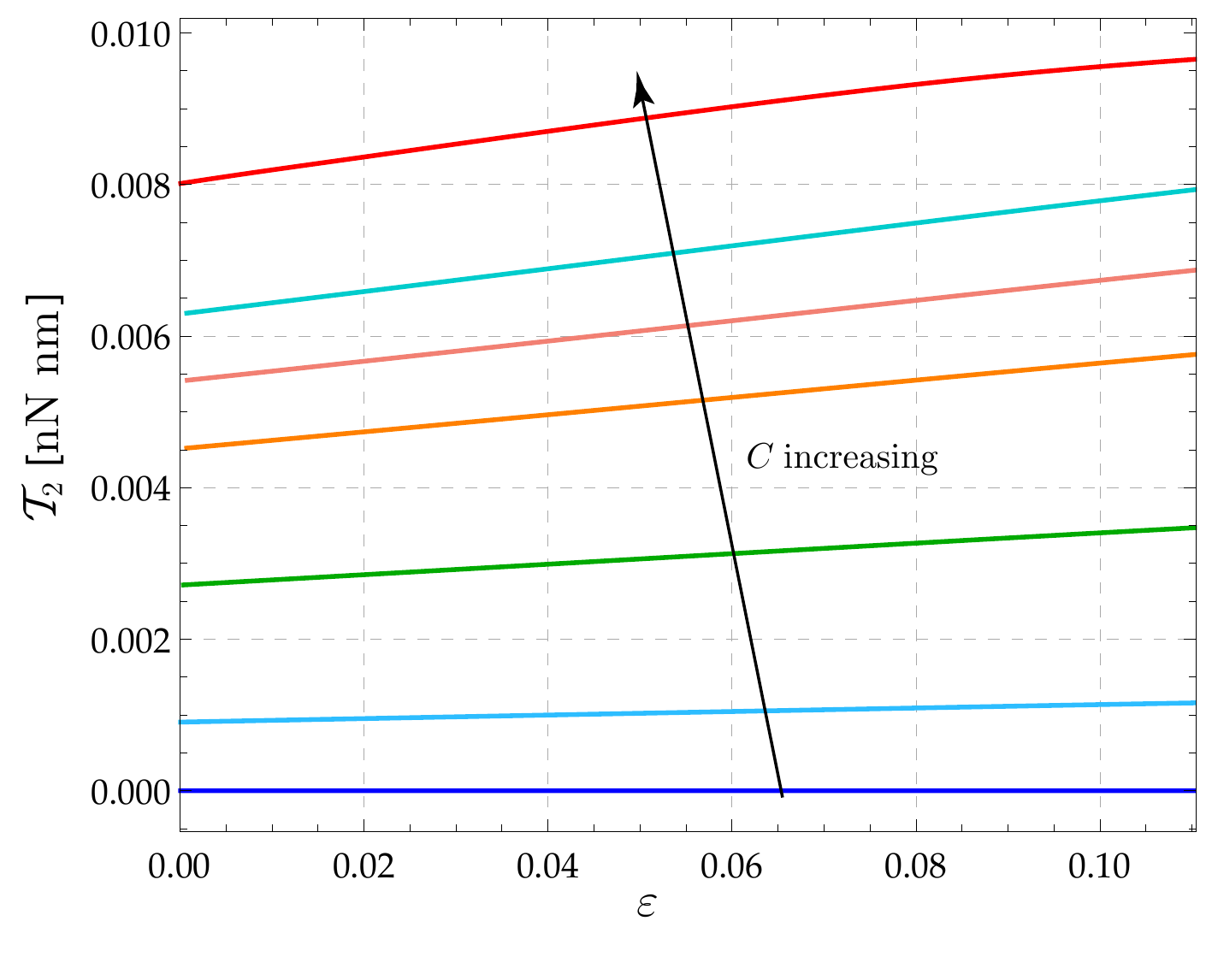}
																										\caption{Nanostress $\Tc_2$ versus  axial strain $\varepsilon$, armchair (left)  and zigzag (right).   }
																										\label{T2}
																									\end{figure}
																									\begin{figure}[h]
																										\centering
																										\includegraphics[scale=0.5]{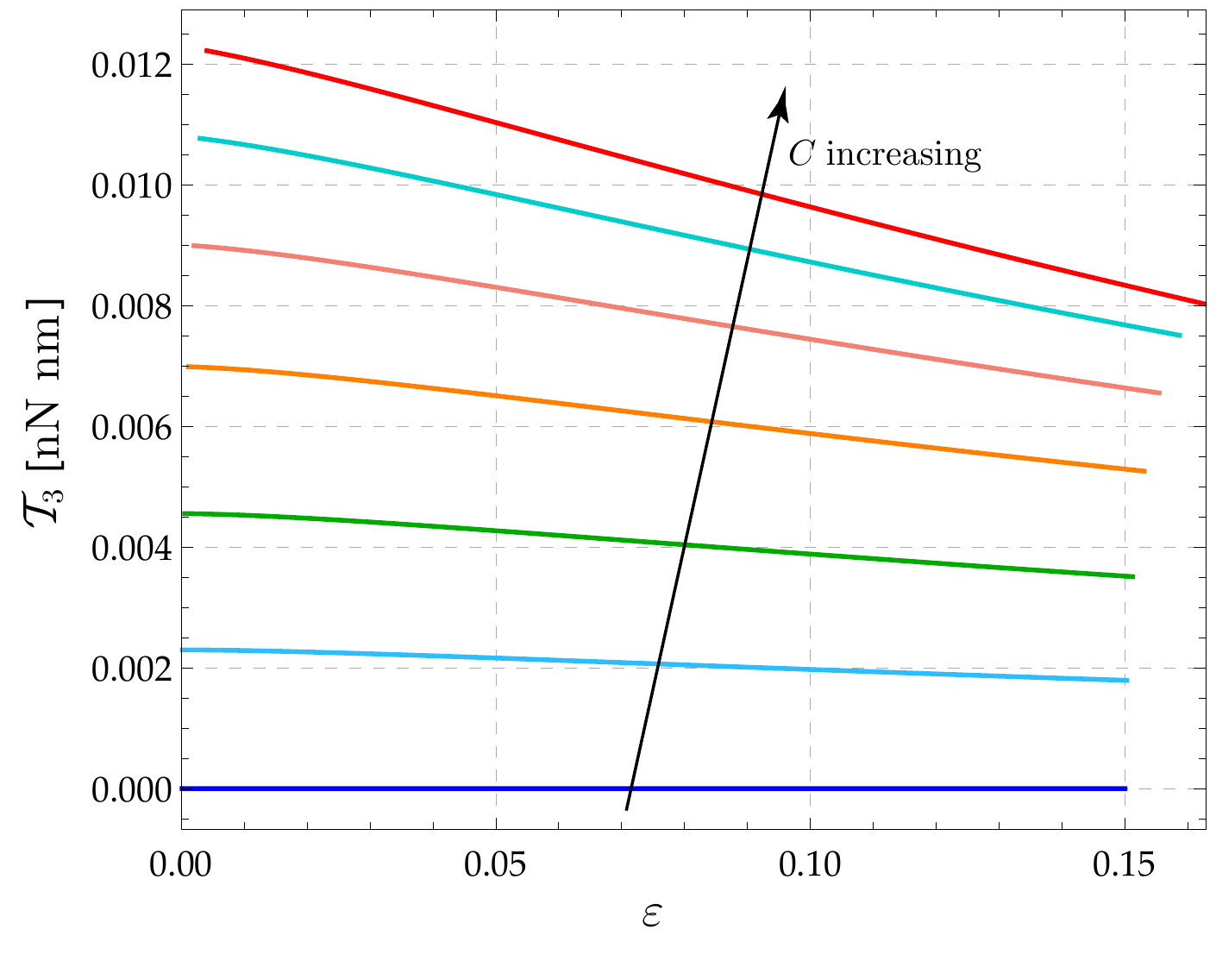}
																										\quad 	\includegraphics[scale=0.5]{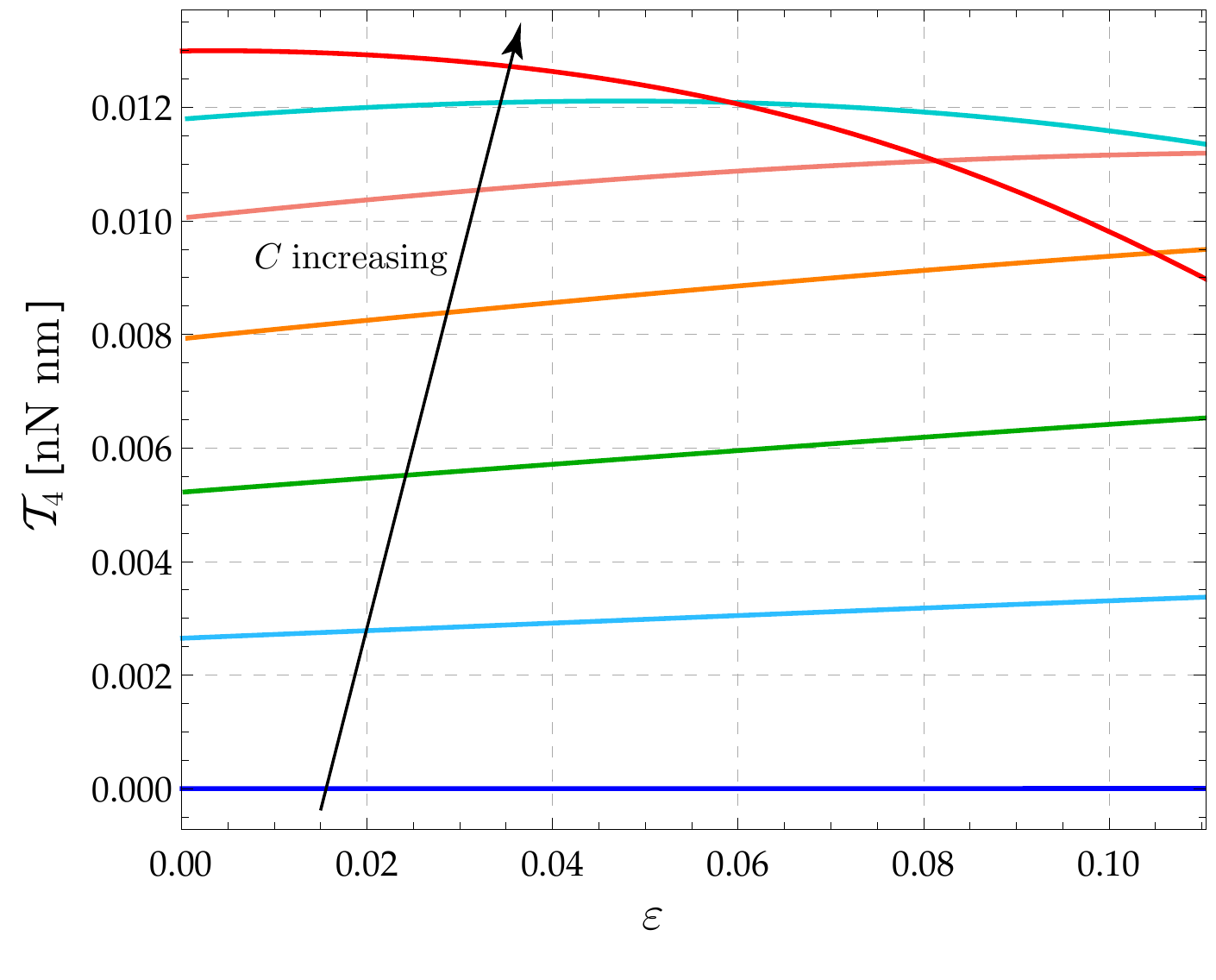}
																										\caption{Nanostress $\Tc_3$ (armchair, left)  and $\Tc_4$ (zigzag, left) versus  axial strain $\varepsilon$.   }
																										\label{T3}
																									\end{figure}
																									illustrate the significant dependence of dihedral-angle nanostresses on the applied couple.

																									\newpage
																									\section{Conclusions}
																									
																									A discrete mechanical model for graphene,  both geometrically and physically nonlinear, has been proposed. Atomic interactions have been specified by choosing a class of REBO potentials depending on strings of kinematic descriptors($\equiv$ order-parameters) identified with changes in bond lengths, bond angles, and dihedral angles. The  equilibrium problem considered has been that of balanced and uniform boundary distributions of force and couple over  pairs of opposite sides of a rectangular graphene sheet. The governing equilibrium equations have been written  in terms of nanostresses, i.e., force-like objects  in one-to-one  correspondence with the order parameters. Suitable definitions of bending and stretching stiffnesses have been proposed, and  analytical formulas given to evaluate them whatever the loads and the configuration, including the ground one; moreover, notions of sensitivity to  changes in applied forces and couples of bending and stretching stiffnesses have been introduced, and two analytical conditions  for detecting softening of the former  and hardening of the latter have been formulated.
																									Such definitions and conditions are written in terms of bond-length, bond-angle, and dihedral, stiffnesses and of bond-angle selfstress, that is, in terms of the quantities on which, according to our discrete model, graphene's mechanical response depends, and whose role, so we believe, should be properly reflected into whatever homogenization procedure one may think of. 
																									
																									Combination of large  bending and stretching states has been investigated here for the first time. It has been shown that concomitant bending and stretching, whatever their value,  concur to make bending stiffness decrease. It has also been shown that concomitant bending and stretching make the stretching stiffness increase until the applied forces reach a threshold value, then they make  it  decrease; the reasons of this rather surprising behaviour have been discussed. Finally, equilibrium nanostresses have been quantitatively evaluated for whatever combination of applied forces and couples.

																								}
																								
																								\addcontentsline{toc}{section}{Acknowledgements}
																								\section*{Acknowledgements}
																								N.M.P. is supported by the European Research Council (ERC StG Ideas 2011  no. 279985, ERC PoC 2015  No. 693670 ,  ERC PoC 2013-2  No. 632277), by the European Commission under the Graphene Flagship (WP10 ``Nanocomposites'', No. 604391).

\end{document}